\documentclass{article}
\usepackage{amsthm, amssymb, amsmath}
\usepackage{enumerate}
\usepackage{graphicx}
\usepackage{xcolor}
\usepackage{braket}
\numberwithin{equation}{section}
\newtheorem{theorem}{Theorem}[section]

\newtheorem{lemma}[theorem]{Lemma}

\newtheorem{proposition}[theorem]{Proposition}

\huge

\begin{document}
\title{ Entropic quantum error correction for independent and correlated amplitude damping errors}
\author{Zhou Gang\footnote{gangzhou@binghamton.edu, partly supported by Simons travel support 709542.}}
\maketitle
\centerline{Department of Mathematics and Statistics, Binghamton University, Binghamton, NY, 13902
}
\setlength{\leftmargin}{.1in}
\setlength{\rightmargin}{.1in}
\normalsize \vskip.1in
\setcounter{page}{1} \setlength{\leftmargin}{.1in}
\setlength{\rightmargin}{.1in}
\large

\date

%\fixNumberingInArticle
\setlength{\leftmargin}{.1in}
\setlength{\rightmargin}{.1in}
%\NowFootNum \fixNumberingInArticle
\normalsize \vskip.1in
\setcounter{page}{1} \setlength{\leftmargin}{.1in}
\setlength{\rightmargin}{.1in}
\large

\section*{Abstract}
The idea of using quantum entropy for quantum error correction (QEC) has existed for a long time. However, its computational complexity has limited its practical use: even a $[[4,1]]$ code for standard independent amplitude damping errors involves a $256\times 256$ Hermitian matrix, so the approach scales poorly as the codeword length increases. For correlated errors, the situation is even worse. In this paper, we show that, by using the insights developed in our previous paper and exploiting certain structures of quantum conditional mutual information, we can solve these problems exactly by hand, using only $2\times 2$ unitary matrices. By examining the entropy, we identify the leakage channels and the amount of leakage exactly, then define the recovery map accordingly. The rules defining our recovery map are simple, intuitive, and computationally inexpensive: for independent errors, it is based on 16 explicitly given orthonormal vectors and 6 instructions; for correlated errors, it is based on 6 explicitly given orthonormal vectors and 6 instructions. Our method sacrifices little in terms of performance: for the widely used entanglement fidelity, our simple QEC recovery map achieves a sharp result; for the worst-case fidelity, the best result is about $9\%$ better than ours; for correlated errors, there are no comparable results for short codewords.
Among the known techniques, our recovery map is the first non-global one.
The calculations use elementary linear algebra and calculus, and is presented in great detail.
\tableofcontents

\section{Introduction}
Developing scalable quantum error correction (QEC) is a major bottleneck in achieving practical fault-tolerant quantum computing; see, for example, \cite{FowlerMartinis2012}.

In this paper, we study entropy-based QEC for both independent and correlated amplitude damping errors. Our goal is to develop a non-global recovery map based on quantum entropy with little loss in performance, in the hope of improving scalability.

We begin with a brief review of existing results.

The Knill--Laflamme condition \cite{KnillLaflamme2000} assumes that errors map the code subspace into nearly orthogonal error subspaces, allowing exact recovery. Thus, it does not work well for amplitude damping errors.

The study of approximate quantum error correction began with \cite{LeungEtc1997ApproQEC}.

For independent amplitude damping errors and short codewords, Fletcher, Shor, and collaborators used semidefinite programming in \cite{FletcherShor2007Semidefinite, fletcher2008channel}, while Biswas and Mandayam used the Petz map in \cite{BiswasEtc2024, biswas2025universal}. However, these techniques are global because they must account for every error, which makes scaling to long codewords difficult. For example, the standard semidefinite program cannot be extended to a distance-7 code; see, for example, \cite{kosut2009quantum}.

The challenge is greater for correlated errors because the complexity of global recovery increases dramatically. Pauli twirling is used to approximate these errors by standard Pauli errors. However, this approach requires very long codewords to obtain meaningful results because Pauli twirling cannot eliminate correlations between errors; see \cite{brillant2026noise}. It can also overestimate the logical error rate; see \cite{katabarwa2015logical}.

We now discuss entropic quantum error correction, which is the main technique used in this paper.

The entropy involved is quantum conditional mutual information (QCMI). For a tripartite quantum density matrix $\rho_{ABC}$ describing subsystems $A$, $B$, and $C$, the QCMI is defined as
\begin{align}\label{def:mutual1}
\begin{split}
I(A : C|B) &= \operatorname{Tr} \rho_{ABC}(t) \log_2 \rho_{ABC}(t) - \operatorname{Tr} \rho_{AB}(t) \log_2 \rho_{AB}(t)\\ 
    &+ \operatorname{Tr} \rho_B(t) \log_2 \rho_B(t) - \operatorname{Tr} \rho_{BC}(t) \log_2 \rho_{BC}(t)
\end{split}
\end{align} where $\rho_{ABC}$ is the density matrix of the whole system, and $\rho_{AB}(t)$, $\rho_{BC}(t)$, and $\rho_B(t)$ are the reduced density matrices defined by
\begin{align*}
    \rho_{AB}(t) &:= \operatorname{Tr}_C \rho_{ABC}(t), \qquad \rho_{BC}(t) = \operatorname{Tr}_A \rho_{ABC}(t), \qquad
    \rho_B(t)    := \operatorname{Tr}_{AC} \rho_{ABC}(t).
\end{align*} Here, $\operatorname{Tr}_{D}$ denotes the partial trace over subsystem $D$.

In 1973, Lieb and Ruskai proved in \cite{LiebRuskai1973} that
\begin{align}
    I(A:C|B)\geq 0.
\end{align} Petz proved in \cite{Petz1986, Petz1988} that perfect recovery is possible if and only if $I(A:C|B)=0$.

When $I(A:C|B)\gneq 0$, one needs to find a recovery map for QEC. A local recovery map is preferable because it is scalable and is therefore highly sought after. This is the focus of the present paper and subsequent papers.

The idea of using QCMI for quantum error correction dates back to at least 1996. Schumacher and Nielsen began to develop ideas for QEC based on QCMI; see \cite{Schumacher1996, SchumacherNielsen1996, schumacher2002approximate} and the more recent results in \cite{GilyenLloyd2022, FawziRenner2015, kato2019quantum}. This approach connects recoverability with an information-theoretic measure of correlations. However, the problem is difficult because it involves large Choi matrices. For example, for four physical qubits subject to amplitude damping errors, the matrix is $256 \times 256$.

In the recent papers \cite{GilyenLloyd2022, FawziRenner2015, kato2019quantum, JuneRenner2018}, recovery maps are constructed using entropies. However,
the gate count scales exponentially with the code length, partly because the operators involved require manipulating the entire density matrix. This makes implementation difficult; for example, implementing the inverse of a large density matrix requires a deep circuit.

New ideas for recovery maps were developed in \cite{Ng2010, biswas2025universal}. However, these techniques are still global, and only independent errors were considered.

We now present our results.

In the present paper, we use the four-qubit Leung code \cite{LeungEtc1997ApproQEC} and follow the line of reasoning in \cite{Schumacher1996, SchumacherNielsen1996, schumacher2002approximate}. The system has three parts: the reference $R$, the logical state system $S$, and the environment $E$.
The QCMI in \eqref{def:mutual1} then takes the form
\begin{align}\label{def:Mutual}
\begin{split}
    I(R : E \mid S) :=& \operatorname{Tr} \rho_{RSE}(t) \log_2 \rho_{RSE}(t) - \operatorname{Tr} \rho_{RS}(t) \log_2 \rho_{RS}(t)\\ 
    &+ \operatorname{Tr} \rho_S(t) \log_2 \rho_S(t) - \operatorname{Tr} \rho_{SE}(t) \log_2 \rho_{SE}(t).
\end{split}
\end{align} Here, the reference system is not subject to any external forces, $S$ is the codeword system, and $E$ is the environment. Even though the codeword length is only 4, the codeword system and the environment involve a $256\times 256$ Choi matrix, which is difficult to process.

In the present paper, we solve the problem by hand in a reasonable amount of time, using only $2\times 2$ unitary matrices. Building on the insights developed in our previous paper \cite{GZ2026exact} and exploiting the structure of the entropy, we let the entropy dictate most of the recovery rules, with the help of three technical lemmas: Lemmas \ref{LM:ChangeBasis}, \ref{LM:simplification}, and \ref{LM:Largest}. By examining the list of entropy leakages, we identify exactly the leakage channels, the amount of leakage, and the best way to correct the errors. We then define the recovery map accordingly.

We also extend our technique to correlated errors and identify the structure of their correlations.

Our QEC technique is non-global because the entropy provides a list of leakage errors ordered by the amount of leakage, and we construct the recovery map to correct the largest leakages and ignore the rest. This differs from the existing techniques in \cite{FletcherShor2007Semidefinite, GilyenLloyd2022, FawziRenner2015, kato2019quantum, Ng2010, biswas2025universal}.

Here we can not claim that our recovery map is local because the codeword considered here is too short.

The problem of constructing local recovery maps for various error models will be addressed in future work, including models of ZZ crosstalk \cite{paetznick2026improved, zhou2025surface, gicev2026crosstalk}. As discussed earlier, an efficient local recovery map is desirable because it holds the promise of overcoming scalability limitations on the path to practical fault-tolerant quantum computation.

Next, we discuss the performance of our recovery maps.

Our instructions for implementing the recovery maps are simple and intuitive. They are given in \eqref{eq:rules-11}--\eqref{eq:r4} for independent errors and in \eqref{eq:corrRule1}--\eqref{eq:corrRule2} for correlated errors. For independent errors, the recovery map is based on 16 explicitly given orthonormal vectors and 6 instructions, listed in \eqref{def:base1}-\eqref{def:base3} and \eqref{eq:rules-11}-\eqref{eq:r4}, respectively. For correlated errors, it is based on 6 explicitly given orthonormal vectors in \eqref{eq:corrBase} and 6 instructions in \eqref{eq:corrRule1} and \eqref{eq:corrRule2}.

To quantify performance, we use the widely adopted entanglement fidelity, denoted by $F_{\text{Ent}}$ in \eqref{eq:biswas} and \eqref{eq:GZ} below; see \cite{nielsen1996entanglement}.
Our simple recovery map achieves a sharp result for this metric, matching the result obtained in \cite{FletcherShor2007Semidefinite} using a method designed to optimize entanglement fidelity.

To take more metrics into consideration, the following results were obtained in \cite{biswas2025universal}, where the parameter $\gamma$ corresponds to our $t$:
\begin{align}\label{eq:biswas}
\begin{split}
F_{\text{Ent}} &= 1 - 1.25 \gamma^2 + O(\gamma^3), \\
F_{\text{min}}^2 &= 1 - 1.15 \gamma^2 + O(\gamma^3)
\end{split}
\end{align} where $F_{\text{Ent}}$ is the entanglement fidelity and $F_{\min}^{2}$ is the worst-case fidelity.
For independent errors, our Theorems \ref{thm:entangleFidelity} and \ref{thm:worstgeneral} below establish the following results for these two metrics:
\begin{align}\label{eq:GZ}
    F_{\text{Ent}}=F_{\min}^{2}=1-1.25 t^2+O(t^3).
\end{align}
Thus, our method achieves the same sharp result for entanglement fidelity, while our result for worst-case fidelity is about $9\%$ worse. The performance of other recovery maps is reported in \cite{biswas2025universal}. As discussed earlier, our method and the associated computations are much simpler, with little loss in performance. Moreover, our technique is non-global.

For correlated errors, the limitations of existing techniques require very long codewords to obtain meaningful results. Since our codeword is short, there are no comparable results available.

The paper is organized as follows. In Sections \ref{sec:mainInde} and \ref{sec:mainCorre}, we formulate the main results for independent and correlated errors. In Section \ref{sec:QuanMutualInfo}, we develop a convenient expression for quantum conditional mutual information. In Sections \ref{sec:IndepEntropy}-\ref{sec:CorrWorstFideltiy}, we prove the results stated in Sections \ref{sec:mainInde} and \ref{sec:mainCorre}. In Appendices \ref{app:orthogonalization}, \ref{app:leakage}, and \ref{app:convex}, we provide technical results.

\section{Main results for independent amplitude damping errors}\label{sec:mainInde}
We begin by defining the quantum system, which consists of three subsystems: the reference system, denoted by $R$; the logical state system, denoted by $S$; and the radiation field, denoted by $E$.

Initially, the system is in the pure
state
\begin{align}\label{eq:initial}
\ket{V_{\mathrm{ini}}}
:=\frac{1}{\sqrt{2}}\ket{R_1}\ket{0_L}\ket{\Omega}
+\frac{1}{\sqrt{2}}\ket{R_2}\ket{1_{L}}\ket{\Omega},
\end{align}
where $\ket{R_1}$ and $\ket{R_2}$ are reference states that are not
subject to any external forces, while $\ket{0_L}$ and $\ket{1_{L}}$ are
the logical states.
Here, we consider the Leung code \cite{LeungEtc1997ApproQEC}, whose logical states $\ket{0_L}$ and $\ket{1_{L}}$ are defined in terms of four physical qubits:
\begin{align}\label{def:Leung}
\begin{split}
    \ket{0_L} :=& \frac{1}{\sqrt{2}}\big(\ket{0000} + \ket{1111}\big),\\ 
    \ket{1_{L}} :=& \frac{1}{\sqrt{2}}\big(\ket{0011} + \ket{1100}\big).
\end{split}
\end{align} For each physical qubit, the ground state is $\ket{0}$ and the excited state is $\ket{1}$. These states are represented by the two-dimensional vectors
\begin{align}\label{def:GSES}
    \ket{0} := 
    \begin{pmatrix} 
    1 \\ 
    0 
    \end{pmatrix};\ \text{and}\ 
    \ket{1} := 
    \begin{pmatrix} 
    0 \\ 
    1 
    \end{pmatrix}
\end{align}

The state $\ket{\Omega}$ is the vacuum state of the radiation field.

For the quantum model governing the evolution of the whole system, we use the standard Hamiltonian together with ideas from \cite{frohlich2025theory, frohlich2024tentative}.
We assume that the speed of light is $c=\infty$. This assumption is reflected in the definition of the operator $P_k$ in \eqref{def:Hamil} and in its property stated below.
The sole purpose of this assumption is to make the present paper fully mathematically rigorous
without discussing the Born--Oppenheimer approximation. We emphasize
that the Kraus operators derived here are the same as those in the existing
literature, so this assumption does not affect the results. Moreover, the assumption $c=\infty$ is consistent with current
laboratory technology.

The Hamiltonian for the whole system is
\begin{align}\label{def:Hamil}
H = Id_{R}\otimes H_{SE} =:Id_{R}\otimes \sum_{k=1}^{4}\big(Id_S \otimes P_{k} + H_{k}\big)
\end{align}
where $H_{SE}$ is the Hamiltonian acting on the logical state system and the environment. Each term $Id_S \otimes P_{k} + H_{k}$ acts only on the $k$-th qubit and its environment. We assume that the actions on different qubits are identical and independent. Here,
$$H_{k}:=
\sigma_k^{+} \otimes a_k^*(0) +  
\sigma_k^{-} \otimes a_k(0)$$
and 
$$
\sigma_k^{+}:=
\begin{bmatrix}
0&1\\
0&0
\end{bmatrix},\  \sigma_k^{-}:=
\begin{bmatrix}
0&0\\
1&0
\end{bmatrix}$$
are the raising and lowering operators on the $k$-th qubit.
The operator $P_k$ is the Hamiltonian of the radiation field and satisfies
\begin{align}\label{eqn:timeTran}
e^{itP_k} a_k^\#(s) e^{-itP_k} = a_k^\#(t+s)
\end{align}
The creation and annihilation operators $a_k^*(t)$ and $a_k(t)$ satisfy
\begin{align}\label{eq:commute}
\begin{split}
[a_k(t), a_k^*(s)] = \delta(t-s), &\quad [a_k^\#(t), a_k^\#(s)] = 0;  \\
[a_\ell^{\#_1}(t), a_k^{\#_2}(s)] = 0 &\text{ when } k \neq \ell,
\end{split}
\end{align}
where $a_{\ell}^{\#} = a_\ell \text{ or } a_\ell^*,$ $l=1,\cdots,4.$

The rest of this section is divided into two parts. In the first part, we identify the leakage channels and the amount of leakage by studying the density matrices and the entropy. In the second part, which begins after \eqref{eq:indAmount}, we use this information to construct a recovery map to correct the errors.

We now consider the density matrix
\begin{align}
\rho_{RSE}:=e^{-itH}\ket{V_{\mathrm{ini}}}\bra{V_{\mathrm{ini}}}e^{itH},
\end{align} where the initial state $\ket{V_{\mathrm{ini}}}$ is defined in \eqref{eq:initial}.
Before stating the result, we define the following vectors.
The vectors $\ket{S_{k,1}}$, $k=1,2,\ldots,11$, are generated by the component corresponding to the logical state $\ket{0_L}$:
\begin{align}\label{def:sk1}
\begin{array}{ll}
\ket{S_{1,1}}:=
\ket{0000}+(1-2t+2t^2)\ket{1111},\ & \
\ket{S_{2,1}}:=\sqrt{t(1-\frac{3}{2}t)} (1-t) \ket{1110},\\
\ket{S_{3,1}}:= \sqrt{t(1-\frac{3}{2}t)} (1-t)\ket{1101},\ & \ 
\ket{S_{4,1}}:= \sqrt{t(1-\frac{3}{2}t)} (1-t)\ket{1011},\\
\ket{S_{5,1}}:= \sqrt{t(1-\frac{3}{2}t)}  (1-t)\ket{0111},\ & \ 
\ket{S_{6,1}}:= \frac{t}{\sqrt{2}}
\left(\ket{1100}+\ket{0011}\right),\\
\ket{S_{7,1}}:=\frac{t}{\sqrt{2}}
\left(\ket{0011}-\ket{1100}\right),\ & \
\ket{S_{8,1}}:= t \ket{1010},\\
\ket{S_{9,1}}:= t \ket{0110},\ & \
\ket{S_{10,1}}:= t \ket{1001},\\
\ket{S_{11,1}}:= t \ket{0101}. & 
\end{array}
\end{align}
The vectors $\ket{S_{k,2}}$, $k=1,2,\ldots,6$, are generated by the component corresponding to the logical state $\ket{1_L}$:
\begin{align}\label{def:sk2}
\begin{array}{ll}
\ket{S_{1,2}}:=(1-t+\frac{1}{2}t^2)
\left(
\ket{1100}+\ket{0011}
\right),\ &\ 
\ket{S_{2,2}}:=\sqrt{t(1-\frac{3}{2}t)}  \ket{0010},\\
\ket{S_{3,2}}:= \sqrt{t(1-\frac{3}{2}t)} \ket{0001},\ & \ 
\ket{S_{4,2}}:=\sqrt{t(1-\frac{3}{2}t)} \ket{1000},\\
\ket{S_{5,2}}:= \sqrt{t(1-\frac{3}{2}t)} \ket{0100},\ & \
\ket{S_{6,2}}:=t \sqrt{ 2 } \ket{0000}.
\end{array}
\end{align}

In Section \ref{sec:IndepDensity}, specifically in Lemma \ref{LM:density}, we prove that the density matrix takes the form
\begin{align}
\rho_{RS}=\operatorname{Tr}_E \rho_{RSE}={}&\frac{1}{4}\sum_{k=1}^{6}
\Big(\ket{R_1}\ket{S_{k,1}}+\ket{R_2} \ket{S_{k,2}}\Big)
\Big(\bra{S_{k,1}}\bra{R_1}+\bra{S_{k,2}}\bra{R_2}\Big)\nonumber\\
&\quad+\frac{1}{4}\sum_{\ell=7}^{11}
\ket{R_1}\ket{S_{\ell,1}}
\bra{S_{\ell,1}}\bra{R_1}+O(t^3).\label{eq:updatedRhoRS}
\end{align}

Two important sources of difficulty are that 
\begin{align}\label{eq:adverseOverlap}
\ket{S_{1,1}}\not\perp \ket{S_{6,2}},\ \text{and}\ \ket{S_{1,2}}\parallel \ket{S_{6,1}}.
\end{align}
These properties cause what we call entropy leakage, which is the main difficulty in our analysis.

In Section \ref{sec:IndepEntropy}, we prove that the QCMI defined in \eqref{def:Mutual} satisfies
\begin{align}\label{eq:IndepEntropy}
    I(R : E \mid S) \geq \frac{1}{4}\operatorname{Tr}_S(D_1+D_2)+O(t^3 \ln t)
\end{align} where $D_1$ and $D_2$ are self-adjoint operators defined as
\begin{align*}
    D_1:=&-\ket{S_{1,1}}\bra{S_{1,1}}
    \log_2\left(\frac{1}{2}\ket{S_{1,1}}\bra{S_{1,1}}\right)-\ket{S_{6,2}}\bra{S_{6,2}}
\log_2\left(\frac{1}{2}\ket{S_{6,2}}\bra{S_{6,2}}\right)\\
&+\Big(
\ket{S_{1,1}}\bra{S_{1,1}}
+\ket{S_{6,2}}\bra{S_{6,2}}
\Big)\log_2\left(
\frac{1}{2}\Big(
\ket{S_{1,1}}\bra{S_{1,1}}
+\ket{S_{6,2}}\bra{S_{6,2}}
\Big)
\right)
\end{align*}
\begin{align*}
D_2:= 
&-\ket{S_{6,1}}\bra{S_{6,1}}
\log_2\left(\frac{1}{2}\ket{S_{6,1}}\bra{S_{6,1}}\right)-\ket{S_{1,2}}\bra{S_{1,2}}
\log_2\left(\frac{1}{2}\ket{S_{1,2}}\bra{S_{1,2}}\right)\\
&+\Big(
\ket{S_{1,2}}\bra{S_{1,2}}
+\ket{S_{6,1}}\bra{S_{6,1}}
\Big)\log_2\left(
\frac{1}{2}\Big(
\ket{S_{1,2}}\bra{S_{1,2}}
+\ket{S_{6,1}}\bra{S_{6,1}}
\Big)
\right).
\end{align*}

For the amount of leakage caused by the space overlap, we use Lemma \ref{LM:simplification} to find, for some positive constants $b_1$ and $b_2$,
\begin{align}\label{eq:indAmount}
\begin{split}
\operatorname{Tr}_SD_1=&b_1 t^2 |\log_2 t|+O(t^3\ln t),\\
\operatorname{Tr}_SD_2=&b_2 t^2 |\log_2 t|+O(t^3\ln t).
\end{split}
\end{align}

We have now identified the leakage channels and the amount. This completes the first part of our analysis.

In the second part, we construct a recovery map to correct the errors. To do so, we need to study
the overlap between the supports of the following operators in more detail:
$
\ket{S_{1,1}}\bra{S_{1,1}}
$ and $
\ket{S_{6,2}}\bra{S_{6,2}}.$
Since $\ket{S_{1,2}}\parallel\ket{S_{6,1}}$, we do not need to analyze the overlap between these two states further.

Next, we apply Lemma \ref{LM:ChangeBasis} to find an orthonormal basis $\{\ket{X_1},\ket{X_3}\}$ for
the subspace $\operatorname{span}\{\ket{S_{1,1}},\ket{S_{6,2}}\}$ such that
in this basis, the matrix representing
$
\ket{S_{1,1}}\bra{S_{1,1}}+\ket{S_{6,2}}\bra{S_{6,2}}
$
is diagonal.

As a preliminary step, we define an orthonormal basis $\{ \ket{Z_1}, \ket{Z_2}\}$ for $\operatorname{span}\{\ket{S_{1,1}},\ket{S_{6,2}}\}$ by
\begin{align}\label{eq:z12}
\begin{bmatrix}
\ket{Z_1}\\[6pt]
\ket{Z_2}
\end{bmatrix}:=\frac{1}{\sqrt{1+(1-\kappa(t))^2}}
\begin{bmatrix}
1 & 1-\kappa(t)\\[6pt]
1-\kappa(t) & -1
\end{bmatrix}
\begin{bmatrix}
\ket{0000}\\[6pt]
\ket{1111}
\end{bmatrix},
\end{align} where $\kappa$ is defined by 
\begin{align}
\kappa(t):=&2t-2t^2.\label{def:kappat}
\end{align}
These definitions yield
\begin{align}\label{eq:s11Norm}
\| S_{1,1}\rangle\|
&=\sqrt{1+(1-\kappa(t))^2},
\end{align}
\begin{align}\label{eq:norS11}
\frac{1}{\| S_{1,1}\rangle\|}
\begin{bmatrix}
\ket{S_{1,1}}\\[6pt]
\ket{S_{6,2}}
\end{bmatrix}
&=
\begin{bmatrix}
1 & 0\\[6pt]
\alpha_1 & \beta_1
\end{bmatrix}
\begin{bmatrix}
\ket{Z_1}\\[6pt]
\ket{Z_2}
\end{bmatrix}.
\end{align}
Using these expressions, we obtain the parameters needed in Lemma \ref{LM:ChangeBasis}:
\begin{align*}
\theta_1
&:=-\frac{1}{2}\arcsin\left(
\frac{2\alpha_1\beta_1}
{\sqrt{(1+\alpha_1^2-\beta_1^2)^2+4\alpha_1^2\beta_1^2}}
\right),\\
\alpha_1
&:=\frac{\sqrt{2}\,t}{1+(1-\kappa(t))^2},
\qquad
\beta_1:=(1-\kappa(t))\alpha_1,
\end{align*}

Applying Lemma \ref{LM:ChangeBasis}, we obtain the desired new basis:
\begin{align}\label{def:x1x3}
\begin{bmatrix}
\ket{X_1}\\[6pt]
\ket{X_3}
\end{bmatrix}
:=
\begin{bmatrix}
\cos\theta_1 & -\sin\theta_1\\[6pt]
\sin\theta_1 & \cos\theta_1
\end{bmatrix}
\begin{bmatrix}
\ket{Z_1}\\[6pt]
\ket{Z_2}
\end{bmatrix}.
\end{align}
The new basis has the desired property because the operator
$\ket{S_{1,1}}\bra{S_{1,1}}+\ket{S_{6,2}}\bra{S_{6,2}}$
has a diagonal representation in this basis:
\begin{align}\label{eq:s11s62Diagonal}
&\ket{S_{1,1}}\bra{S_{1,1}}
+\ket{S_{6,2}}\bra{S_{6,2}}\nonumber\\
&\quad=\left(1+(1-\kappa(t))^2\right)
\left[
\frac{1+\alpha_1^2+\beta_1^2+\Delta_1}{2}
\ket{X_1}\bra{X_1}
+\frac{1+\alpha_1^2+\beta_1^2-\Delta_1}{2}
\ket{X_3}\bra{X_3}
\right],
\end{align} where $\Delta_1$ is defined by
$
    \Delta_1:=\sqrt{(1+\alpha_1^2-\beta_1^2)^2+4\alpha_1^2\beta_1^2}.
$

The following identity will be needed later.
\begin{align}\label{eq:zzzz}
\ket{0000}
&=\frac{1}{\sqrt{1+(1-\kappa(t))^2}}
\left[\ket{Z_1}+(1-\kappa(t))\ket{Z_2}\right].
\end{align}

Before defining the recovery map $\mathcal{R}$, we specify the full orthonormal basis.

Recall that $\ket{X_1}$ and $\ket{X_3}$ are defined in \eqref{def:x1x3}. We also define the vector
\begin{align}
\ket{X_2}:=&\frac{1}{\sqrt{2}}
\left(\ket{1100}+\ket{0011}\right)=\ket{1_{L}}.\label{def:X2}
\end{align}

We divide the orthonormal basis into several subsets. 

Define the set $\mathcal{X}$ by
\begin{align}\label{def:base1}
\mathcal{X}:=\left\{\ket{X_1},\ket{X_2},\ket{X_3}\right\}.
\end{align}
For $n=2,3,4,5$, the set $\mathcal{S}_n$ contains two vectors:
\begin{align}\label{def:base2}
\mathcal{S}_n:=\left\{\ket{\widetilde{S}_{n,k}}\;\middle|\;k=1,2\right\},
\end{align}
and for $n=7,8,9,10,11$, the set $\mathcal{S}_n$ contains only one vector:
\begin{align}\label{def:base3}
\mathcal{S}_n:=\left\{\ket{\widetilde{S}_{n,1}}\right\},
\end{align}
where $\ket{\widetilde{S}_{k,l}}$ is the normalized version of $\ket{S_{k,l}}$:
\begin{align*}
\ket{\widetilde{S}_{k,l}}
&:=\frac{1}{\sqrt{\braket{S_{k,l}|S_{k,l}}}}\ket{S_{k,l}}.
\end{align*}
The vectors $\ket{S_{k,1}}$ and $\ket{S_{\ell,2}}$ are defined in
\eqref{def:sk1} and \eqref{def:sk2}, respectively.

We are ready to define the recovery map.

For notational convenience, we define
\begin{align}\label{def:S1S2}
\ket{S_1}:=\ket{0_L},\qquad \ket{S_2}:=\ket{1_L}.
\end{align}
We define the recovery map $\mathcal{R}$ by the following rules:
\begin{itemize}
    \item For the vectors in the set $\mathcal{X}$,
\begin{align}
\mathcal{R}\bigl(\ket{X_k}\bra{X_\ell}\bigr)
&=\ket{S_k}\bra{S_\ell},\qquad k,\ell\in\{1,2\},\label{eq:rules-11}\\
\mathcal{R}\bigl(\ket{X_3}\bra{X_3}\bigr)
&=\ket{1_L}\bra{1_L},\label{eq:rules-12}\\
\mathcal{R}\bigl(\ket{X_3}\bra{X_n}\bigr)
&=\mathcal{R}\bigl(\ket{X_n}\bra{X_3}\bigr)=0,
\qquad n\in\{1,2\}.\label{eq:rules-13}
\end{align}
\item
When $n\in\{2,3,4,5\}$ and $k,\ell\in\{1,2\}$, we set
\begin{align}\label{eq:r2}
\mathcal{R}\bigl(\ket{\widetilde{S}_{n,k}}\bra{\widetilde{S}_{n,\ell}}\bigr)
&=\ket{S_k}\bra{S_\ell}.
\end{align}
\item
When $7\leq n\leq 11$, we set
\begin{align}\label{eq:r3}
\mathcal{R}\bigl(\ket{\widetilde{S}_{n,1}}\bra{\widetilde{S}_{n,1}}\bigr)
&=\ket{0_L}\bra{0_L}.
\end{align}
\item For any two vectors $\ket{v}$ and $\ket{w}$ belonging to different subsets of the orthonormal basis, we set
\begin{align}\label{eq:r4}
\mathcal{R}\bigl(\ket{v}\bra{w}\bigr)=0.
\end{align}
\end{itemize}
The following lemma establishes that $\mathcal{R}$ satisfies the requirements for a recovery map:
\begin{lemma}\label{LM:justification}
The mapping $\mathcal{R}$ is completely positive and trace-preserving.
\end{lemma}
This lemma will be proved in Section \ref{sec:recovery}.

Next, we evaluate the performance of our recovery map. Two widely used metrics are entanglement fidelity and
worst-case fidelity; see
\cite{Barnum2002} and
\cite{Schumacher1996, Ng2010, Zheng2024}.

The entanglement fidelity
$F_e$ is defined as
\begin{align}\label{def:entangleFidelity}
F_e&:=\bra{\psi^{RS}}  \mathcal{R}\circ\mathcal{A}
\left(\ket{\psi^{RS}}\bra{\psi^{RS}}\right)  \ket{\psi^{RS}},
\end{align} where
\begin{align*}
\ket{\psi^{RS}}:=\frac{1}{\sqrt{2}}
\left(\ket{R_1}\ket{0_L}+\ket{R_2}\ket{1_L}\right),
\end{align*} and
we use $\mathcal{A}
\left(\ket{\psi^{RS}}\bra{\psi^{RS}}\right)$ to denote the density matrix of the combined $R$ and $S$ subsystems obtained by tracing out the environment:
\begin{align*}
&\mathcal{A}
\left(\ket{\psi^{RS}}\bra{\psi^{RS}}\right)\\
:={}&\operatorname{Tr}_E\Big(\frac{1}{2}e^{-itH}\Big(
\ket{R_1}\ket{0_L}\ket{\Omega}
+\ket{R_2}\ket{1_{L}}\ket{\Omega}
\Big)\Big(
\bra{R_1}\bra{0_L}\bra{\Omega}
+\bra{R_2}\bra{1_L}\bra{\Omega}
\Big)e^{itH}\Big)\\
={}&\rho_{RS}
\end{align*} where $\rho_{RS}$ is given in \eqref{eq:updatedRhoRS}.

In terms of entanglement fidelity, which is perhaps the most widely used metric, see \cite{Schumacher1996, FletcherShor2007Semidefinite}, our recovery map achieves the following sharp result:
\begin{theorem}\label{thm:entangleFidelity}
\begin{align}\label{eq:indeEntanglement}
\begin{split}
F_e&=1-\frac{5}{4} t^2+O(t^3).
\end{split}
\end{align}
\end{theorem}
This theorem will be proved in Section \ref{sec:IndEntangleFidelity}.

Now we turn to the worst-case fidelity, which is defined as the minimum fidelity over all codewords in the code space $\mathcal{C}$.
Specifically, any normalized state $\ket{\Psi}\in\mathcal{C}$
can be written in the following form for some $\alpha,\ \beta\in\mathbb{C}$ satisfying $|\alpha|^2+|\beta|^2=1$:
\begin{equation*}
\ket{\Psi}=\alpha\ket{0_L}+\beta\ket{1_L}.
\end{equation*}

For notational convenience, we write
\begin{equation*}
\ket{\Psi_{\alpha,\beta}}:=\alpha\ket{0_L}+\beta\ket{1_L}.
\end{equation*}
Recall $H_{SE}$ from \eqref{def:Hamil}. Define
\begin{align*}
\mathcal{A}\left(\ket{\Psi_{\alpha,\beta}}\bra{\Psi_{\alpha,\beta}}\right)
:={}&\operatorname{Tr}_E\Bigl[
\left(e^{-itH_{SE}}\ket{\Psi_{\alpha,\beta}}\ket{\Omega}\right)
\left(\bra{\Psi_{\alpha,\beta}}\bra{\Omega}e^{itH_{SE}}\right)
\Bigr],
\end{align*}
and the worst-case fidelity is defined by
\begin{align}\label{def:worstFidelity}
F_{\min}^{2}(\mathcal{R}\circ\mathcal{A})
&:=\min_{\alpha, \beta\in \mathbb{C}, \ |\alpha|^2+|\beta|^2= 1}
\bra{\Psi_{\alpha,\beta}}\mathcal{R}\circ\mathcal{A}
\bigl(\ket{\Psi_{\alpha,\beta}}\bra{\Psi_{\alpha,\beta}}\bigr)\ket{\Psi_{\alpha,\beta}},
\end{align}
where $\mathcal{R}$ is the recovery map defined in \eqref{eq:rules-11}--\eqref{eq:r4}.

We obtain the following result:
\begin{theorem}\label{thm:worstgeneral}
The worst-case fidelity is
\begin{align}\label{eq:worstFidelity}
F_{\min}^{2}(\mathcal{R}\circ\mathcal{A})=1-\frac{5}{4} t^2+O(t^3).
\end{align}
\end{theorem}
This theorem will be proved in Section \ref{sec:IndWorstFidelity}.

\section{Main results for correlated amplitude damping errors}\label{sec:mainCorre}
We begin by defining the quantum system.

The Hamiltonian for correlated errors has a form similar to that of the Hamiltonian
defined in \eqref{def:Hamil}, but with a different interaction term:
\begin{align}\label{def:hami2}
\begin{split}
H &= Id_R\otimes\widetilde H_{SE}:=Id_R\otimes\left(
Id_S\otimes\sum_{k=1}^{4}P_k+H_I
\right).
\end{split}
\end{align}
Here, $P_k$ is the same as in \eqref{def:Hamil}, and $H_I$ takes the form

\begin{equation*}
H_I=\sum_k\sum_{j=1}^{4}
\left(
g_{kj}\sigma_-^{(j)}a_k(0)
+g_{kj}^*\sigma_+^{(j)}a_k^*(0)
\right),
\end{equation*}
where $\sigma_+^{(j)}$ and $\sigma_-^{(j)}$ are the raising and lowering
operators for the $j$-th qubit.

As in \eqref{eqn:timeTran},
\begin{align}\label{def:corrAa}
a_k^*(t)=e^{itP}a_k^*(0)e^{-itP}\ \ \text{and}\ \
a_k(t)=e^{itP}a_k(0)e^{-itP}.
\end{align}

To make the noise fully correlated, assume that the spatial separation
between the four qubits is much smaller than the correlation length of
the bath modes. The environment then cannot distinguish the qubits, so
the coupling constants become site-independent:
\begin{equation*}
g_{kj}\approx g_k.
\end{equation*}

We therefore define the collective ladder operators
\begin{align}\label{def:jj}
J_-:=\sum_{j=1}^{4}\sigma_-^{(j)},
\qquad
J_+:=\sum_{j=1}^{4}\sigma_+^{(j)},
\end{align}
and 
\begin{align}\label{def:sumAa}
a(t):=\sum_k g_k a_k(t), \ \ a^{*}(t):=\sum_k\overline{g_k}a_k^{*}(t).
\end{align}
The interaction Hamiltonian can then be written as
\begin{align}\label{eq:CorrHI}
H_I=
J_-a+J_+ a^*
\end{align}
Here, the operators $a^*(t)$ and $a(t)$ satisfy the following commutation relations:
\begin{align*}
[a(t),a^*(s)]&=|g|^2\delta(t-s),\\
[a(t),a(s)]&=[a^*(t),a^*(s)]=0,
\end{align*} where $|g|^2$ is a positive scalar defined by
\begin{align}
    |g|^2:=\sum_{k}|g_{k}|^2.
\end{align}

Because the interaction depends on the collective operator $J_+$, a
single photon loss to the environment does not damp only one qubit. It
creates a coherent superposition of errors across all four qubits, for
example,
\begin{equation*}
\ket{0111}+\ket{1011}+\ket{1101}+\ket{1110}.
\end{equation*}

Before deriving expressions for the density matrices, we define several vectors.

Define
\begin{align}
\begin{split}\label{eq:y1112}
\ket{Y_{1,1}}
&:=\frac{1}{\sqrt{2}}\ket{0000}
+\frac{1}{\sqrt{2}}
\Big(1-2|g|^2t+2|g|^4t^2\Big)\ket{1111},\\
\ket{Y_{1,2}}
&:=\ket{1_{L}}
-\frac{|g|^2}{\sqrt{2}}t
\left(1-\frac{3|g|^2}{2}t\right)\ket{\Psi_3}.
\end{split}
\end{align}
\begin{align}\label{eq:y212231}
\ket{Y_{2,1}}&:=\ket{\Psi_1}, &\qquad
\ket{Y_{2,2}}
&:=\ket{\Psi_2},\\
\ket{Y_{3,1}}
&:= |g|^2 t\ket{\Psi_3},&\qquad
\ket{Y_{3,2}}
&:=2|g|^2 t\ket{0000},\nonumber
\end{align}
where the vectors $\ket{\Psi_k}$, $k=1,2,3$, are defined by
\begin{align}\label{def:Psi123}
\begin{split}
\ket{\Psi_1}:={}&\ket{1110}+\ket{1101}+\ket{1011}+\ket{0111},\\
\ket{\Psi_2}:={}&\ket{1000}+\ket{0100}+\ket{0010}+\ket{0001},\\
\ket{\Psi_3}:={}&\ket{1100}+\ket{1010}+\ket{1001}+\ket{0110}+\ket{0101}+\ket{0011}.
\end{split}
\end{align}
We also define three vectors:
\begin{align}\label{def:CorrW123}
\begin{split}
\ket{W_1} &:= \ket{R_1}\ket{Y_{1,1}} + \ket{R_2}\ket{Y_{1,2}},\\
\ket{W_2} &:= \sqrt{\frac{1}{2}|g|^2 t(1-5|g|^2 t)}\Big(\ket{R_1}\ket{Y_{2,1}} + \ket{R_2}\ket{Y_{2,2}}\Big),\\
\ket{W_3} &:= \ket{R_1}\ket{Y_{3,1}} + \ket{R_2}\ket{Y_{3,2}}.
\end{split}
\end{align}
These three vectors are mutually orthogonal:
\begin{align}\label{eq:CorOrtho}
    \langle W_k | W_{l}\rangle=0,\ \text{if}\ k\not=l.
\end{align}

The following proposition will be proved in Section \ref{sec:CorrProof}.
\begin{proposition}
The density matrix of the system is
\begin{align}\label{eq:CorrRho}
\rho_{RS}(t)
=\frac{1}{2}\sum_{k=1}^{3}
\ket{W_k}\bra{W_k}+O(t^3).
\end{align}
\end{proposition}

In the analysis that follows, we need to account for the following pairs of non-orthogonal vectors:
\begin{align}\label{eq:CorrNonOrtho}
\ket{Y_{1,1}} &\not\perp \ket{Y_{3,2}},\qquad
\ket{Y_{1,2}} \not\perp \ket{Y_{3,1}}.
\end{align}
All other pairs are orthogonal, as desired.

In Section \ref{sec:corrEntropy}, we prove that the quantum conditional mutual information satisfies
\begin{align}\label{eq:corrEntropy}
I(R:E|S)&\geq \frac{1}{2}\operatorname{Tr}_S(E_1+E_2)+O(t^3\ln t),
\end{align} where $E_1$ and $E_2$ are operators defined as
\begin{align*}
E_1
:={}&-\operatorname{Tr}_S\Big(
\ket{Y_{1,1}}\bra{Y_{1,1}}
\log_2\left(\ket{Y_{1,1}}\bra{Y_{1,1}}\right)
\Big)-\operatorname{Tr}_S\Big(
\ket{Y_{3,2}}\bra{Y_{3,2}}
\log_2\left(\ket{Y_{,2}}\bra{Y_{3,2}}\right)
\Big)\\
&+\operatorname{Tr}_S\Big(
\left(
\ket{Y_{1,1}}\bra{Y_{1,1}}
+\ket{Y_{3,2}}\bra{Y_{3,2}}
\right)\ \log_2\left(
\ket{Y_{1,1}}\bra{Y_{1,1}}
+\ket{Y_{3,2}}\bra{Y_{3,2}}
\right)
\Big),\\
E_2
:={}&-\operatorname{Tr}_S\Big(
\ket{Y_{1,2}}\bra{Y_{1,2}}
\log_2\left(\ket{Y_{1,2}}\bra{Y_{1,2}}\right)
\Big)-\operatorname{Tr}_S\Big(
\ket{Y_{3,1}}\bra{Y_{3,1}}
\log_2\left(\ket{Y_{3,1}}\bra{Y_{3,1}}\right)
\Big)\\
&+\operatorname{Tr}_S\Big(
\left(
\ket{Y_{1,2}}\bra{Y_{1,2}}
+\ket{Y_{3,1}}\bra{Y_{3,1}}
\right)\ \log_2\left(
\ket{Y_{1,2}}\bra{Y_{1,2}}
+\ket{Y_{3,1}}\bra{Y_{3,1}}
\right)
\Big).
\end{align*}

By Lemma \ref{LM:simplification}, the overlaps within the two pairs in \eqref{eq:CorrNonOrtho} give rise to the following leakage terms for some positive constants $a_1$ and $a_2$:
\begin{align}
\begin{split}
\operatorname{Tr}_S E_1=&a_1 t^2 |\log_2 t|+O(t^3\ln t),\\
\operatorname{Tr}_S E_2=&a_2 t^2 |\log_2 t|+O(t^3\ln t).
\end{split}
\end{align}

As in the independent case, the overlaps between $\ket{Y_{1,1}}$ and $\ket{Y_{3,2}}$ and between $\ket{Y_{1,2}}$ and $\ket{Y_{3,1}}$ are the dominant cause of entropy leakage. To analyze this leakage, we find an orthonormal basis for the span of each pair in which the corresponding sum of rank-one operators is diagonal. This helps us construct the recovery map.

For the term $E_1$, it is easy to verify that $\{\ket{Z_3},\ket{Z_4}\}$ is an orthonormal basis for
$
\operatorname{span}\{\ket{Y_{1,1}},\ket{Y_{3,2}}\},
$
where, recall the definitions of $\ket{Y_{1,1}}$ and $\ket{Y_{3,2}}$ from \eqref{eq:y212231} and \eqref{eq:y1132}, 
\begin{align*}
\begin{bmatrix}
\ket{Z_3}\\[6pt]
\ket{Z_4}
\end{bmatrix}
&:=\frac{1}{\sqrt{2}}\frac{1}{\| Y_{1,1}\rangle \|}
\begin{bmatrix}
1 & \gamma\\[6pt]
\gamma & -1
\end{bmatrix}
\begin{bmatrix}
\ket{0000}\\[6pt]
\ket{1111}
\end{bmatrix},
\end{align*}
and $\gamma$ is defined by
\begin{equation*}
\gamma:=1-2|g|^2t+2|g|^4t^2.
\end{equation*} These definitions yield
\begin{align}\label{eq:y11Norm}
\|Y_{1,1}\rangle\|=\sqrt{\frac{1+\gamma^2}{2}},
\end{align}
\begin{align}\label{eq:y1132}
\frac{1}{\| Y_{1,1}\rangle \|}
\begin{bmatrix}
\ket{Y_{1,1}}\\[6pt]
\ket{Y_{3,2}}
\end{bmatrix}
=
\begin{bmatrix}
\ket{Z_3}\\[6pt]
\dfrac{2}{\| Y_{1,1}\rangle \|}g^2t\ket{0000}
\end{bmatrix}
=
\begin{bmatrix}
1 & 0\\[6pt]
\dfrac{2\sqrt{2}}{1+\gamma^2}|g|^2t & \dfrac{2\sqrt{2}\gamma}{1+\gamma^2}|g|^2t
\end{bmatrix}
\begin{bmatrix}
\ket{Z_3}\\[6pt]
\ket{Z_4}
\end{bmatrix}.
\end{align}

Thus, the parameters needed to apply Lemma \ref{LM:ChangeBasis} are $\alpha_2$, $\beta_2$, and $\theta_2$, defined by
\begin{equation*}
\alpha_2:=\frac{2\sqrt{2}}{1+\gamma^2}|g|^2t,
\qquad
\beta_2:=\frac{2\sqrt{2}\gamma}{1+\gamma^2}|g|^2t,
\end{equation*}
\begin{align*}
\theta_2
:=-\frac{1}{2}\arcsin{\frac{2\alpha_2\beta_2}
{\sqrt{(1+\alpha_2^2-\beta_2^2)^2+4\alpha_2^2\beta_2^2}}}.
\end{align*}
Applying Lemma \ref{LM:ChangeBasis}, we obtain the desired orthonormal basis from $\ket{Z_3}$ and $\ket{Z_4}$ through a unitary transformation:
\begin{align}\label{def:Phi1Phi2}
\begin{bmatrix}
\ket{W_{1,1}}\\
\ket{W_{3,2}}
\end{bmatrix}
&:=
\begin{bmatrix}
\cos\theta_2 & -\sin\theta_2\\
\sin\theta_2 & \cos\theta_2
\end{bmatrix}
\begin{bmatrix}
\ket{Z_3}\\
\ket{Z_4}
\end{bmatrix}.
\end{align}

This basis has the desired property because
\begin{align}\label{eq:y1142}
&\ket{Y_{1,1}}\bra{Y_{1,1}}+\ket{Y_{3,2}}\bra{Y_{3,2}}\nonumber\\
&\quad=\braket{Y_{1,1}|Y_{1,1}}\Big[
\frac{1+\alpha_2^2+\beta_2^2+\Delta_2}{2}\ket{W_{1,1}}\bra{W_{1,1}}
+\frac{1+\alpha_2^2+\beta_2^2-\Delta_2}{2}\ket{W_{3,2}}\bra{W_{3,2}}
\Big]
\end{align}
with
$
\Delta_2:=\sqrt{(1+\alpha_2^2-\beta_2^2)^2+4\alpha_2^2\beta_2^2}.
$

We now consider the term $E_2$, which we analyze in the same way as $E_1$.

Define the positive parameter $\delta$ by
\begin{equation*}
\delta:=|g|^2t\left(1-\frac{3}{2}|g|^2t\right).
\end{equation*} 
This gives
\begin{align}\label{eq:y12Norm}
\left\|Y_{1,2}\rangle\right\|=\sqrt{(1-\delta)^2+2\delta^2}.
\end{align}

Define two orthonormal vectors $\ket{Z_5}$ and $\ket{Z_6}$ by
\begin{align}
\begin{bmatrix}
\ket{Z_5}\\[6pt]
\ket{Z_6}
\end{bmatrix}
&:=\frac{1}{\sqrt{(1-\delta)^2+2\delta^2}}
\begin{bmatrix}
(1-\delta)\ket{1_L}-\dfrac{\delta}{\sqrt{2}}\ket{\widetilde{\Psi}_3}\\[6pt]
\sqrt{2}\delta\ket{1_L}+\dfrac{1-\delta}{2}\ket{\widetilde{\Psi}_3}
\end{bmatrix}\nonumber\\[6pt]
&=\frac{1}{\sqrt{(1-\delta)^2+2\delta^2}}
\begin{bmatrix}
1-\delta & -\dfrac{\delta}{\sqrt{2}}\\[6pt]
\sqrt{2}\delta & \dfrac{1-\delta}{2}
\end{bmatrix}
\begin{bmatrix}
\ket{1_L}\\[6pt]
\ket{\widetilde{\Psi}_3}
\end{bmatrix},\label{eq:matrixFormZ56}
\end{align}
where $\ket{\widetilde{\Psi}_3}$ is defined as
\begin{equation*}
\ket{\widetilde{\Psi}_3}:=\ket{1010}+\ket{1001}+\ket{0110}+\ket{0101}.
\end{equation*}
By the definition in \eqref{def:Psi123}, we have
$
\ket{\Psi_3}=\ket{\widetilde{\Psi}_3}+\sqrt{2}\ket{1_L},
$
and hence by \eqref{eq:matrixFormZ56},
\begin{align*}
\ket{\Psi_3}
&=\begin{bmatrix}\sqrt{2}, & 1\end{bmatrix}\cdot
\left(
\sqrt{(1-\delta)^2+2\delta^2}
\begin{bmatrix}
1-\delta & -\dfrac{\delta}{\sqrt{2}}\\[6pt]
\sqrt{2}\delta & \dfrac{1-\delta}{2}
\end{bmatrix}^{-1}
\begin{bmatrix}
\ket{Z_5}\\[6pt]
\ket{Z_6}
\end{bmatrix}
\right)\\
&=\frac{1}{\sqrt{(1-\delta)^2+2\delta^2}}
\left[\sqrt{2}(1-3\delta)\ket{Z_5}+2\ket{Z_6}\right],
\end{align*} where $\cdot$ signifies the standard dot product.
This, together with definition of $\ket{Y_{1,2}}$ and $\ket{Y_{3,1}}$, gives
\begin{align}\label{eq:y1231}
\frac{1}{\left\|Y_{1,2}\rangle\right\|}
\begin{bmatrix}
\ket{Y_{1,2}}\\[6pt]
\ket{Y_{3,1}}
\end{bmatrix}
&=
\begin{bmatrix}
1 & 0\\[6pt]
\dfrac{\sqrt{2}(1-3\delta)}{(1-\delta)^2+2\delta^2}g^2t
& \dfrac{2}{(1-\delta)^2+2\delta^2}|g|^2t
\end{bmatrix}
\begin{bmatrix}
\ket{Z_5}\\[6pt]
\ket{Z_6}
\end{bmatrix}.
\end{align}

The parameters needed to apply Lemma \ref{LM:ChangeBasis} are $\alpha_3$, $\beta_3$, and $\theta_3$, defined by
\begin{equation*}
\alpha_3=\frac{\sqrt{2} }{(1-\delta)^2+2\delta^2}|g|^2t(1-3\delta),
\qquad
\beta_3=\frac{2 }{(1-\delta)^2+2\delta^2}|g|^2t,
\end{equation*}
\begin{align*}
\theta_3
:=-\frac{1}{2}\arcsin(\frac{2\alpha_3\beta_3}
{\sqrt{(1+\alpha_3^2-\beta_3^2)^2+4\alpha_3^2\beta_3^2}}),
\end{align*}
and, using the basis formulas in Lemma \ref{LM:ChangeBasis}, we obtain the desired orthonormal basis from $\ket{Z_5}$ and $\ket{Z_6}$ through a unitary transformation:
\begin{align}\label{def:Phi3Phi4}
\begin{bmatrix}
\ket{W_{1,2}}\\
\ket{W_{3,1}}
\end{bmatrix}
&:=
\begin{bmatrix}
\cos\theta_3 & -\sin\theta_3\\
\sin\theta_3 & \cos\theta_3
\end{bmatrix}
\begin{bmatrix}
\ket{Z_5}\\
\ket{Z_6}
\end{bmatrix}.
\end{align}

Then, in the new orthonormal basis, the operator takes the desired diagonal form:
\begin{align}\label{eq:y1241}
&\ket{Y_{1,2}}\bra{Y_{1,2}}+\ket{Y_{3,1}}\bra{Y_{3,1}}\nonumber\\
&\quad=\braket{Y_{1,2}|Y_{1,2}}\Bigg[
\frac{1+\alpha_3^2+\beta_3^2+\Delta_3}{2}\ket{W_{1,2}}\bra{W_{1,2}}
+\frac{1+\alpha_3^2+\beta_3^2-\Delta_3}{2}\ket{W_{3,1}}\bra{W_{3,1}}
\Bigg].
\end{align}
Here, $\Delta_3$ is defined by
$
\Delta_3:=\sqrt{(1+\alpha_3^2-\beta_3^2)^2+4\alpha_3^2\beta_3^2}.
$

We now turn to the definition of the recovery map $\mathcal{R}$.

The orthonormal basis
used here is
\begin{align}\label{eq:corrBase}
\{\ket{W_{k,l}}\},\ k=1,2,3,\ l=1,2,
\end{align}
where $\ket{W_{1,1}}$ and $\ket{W_{3,2}}$ are defined in \eqref{def:Phi1Phi2},
$\ket{W_{1,2}}$ and $\ket{W_{3,1}}$ are defined in \eqref{def:Phi3Phi4}, and $\ket{W_{2,1}}$ and $\ket{W_{2,2}}$ are defined as follows:
\begin{align}\label{def:Phi5Phi6}
\begin{split}
\ket{W_{2,1}}
&:=\frac{1}{\| Y_{2,1}\rangle \|}\ket{Y_{2,1}}
=\frac{1}{2}\ket{\Psi_1},\\
\ket{W_{2,2}}
&:=\frac{1}{\| Y_{2,2}\rangle \|}\ket{Y_{2,2}}
=\frac{1}{2}\ket{\Psi_2}.
\end{split}
\end{align}

We are now ready to specify the rules for the recovery map $\mathcal{R}$. For notational convenience, we define $\ket{S_k}$, $k=1,2$, by
\begin{align}\label{def:s1s2}
\ket{S_1}:=\ket{0_L},
\qquad
\ket{S_2}:=\ket{1_L}.
\end{align}
The recovery map $\mathcal{R}$ is defined by the following rules:
\begin{itemize}
\item 
For $k=1,2,3$ and $n,m=1,2$, we define
\begin{align}\label{eq:corrRule1}
\mathcal{R}\left(\ket{W_{k,n}}\bra{W_{k,m}}\right)
=\ket{S_n}\bra{S_m},
\end{align}
\item
When $k_1\neq k_2$, for $k_1,k_2\in\{1,2,3\}$ and $n,m=1,2$, we set
\begin{align}\label{eq:corrRule2}
\mathcal{R}\left(\ket{W_{k_1,n}}\bra{W_{k_2,m}}\right)=0.
\end{align}
\end{itemize}

The following theorems describe the performance of our recovery map.
\begin{theorem}\label{thm:correlatedEntanglement}
For correlated amplitude damping errors, the entanglement fidelity is
\begin{align}
F_e&
= 1 - \big( 5 - \frac{(1 + \sqrt{2})^2}{2} \big) |g|^4 t^2+O(t^3).
\end{align}
\end{theorem}
This theorem will be proved in Section \ref{sec:CorrEntangleFidelity}.

\begin{theorem}\label{thm:correlatedWorstCase}
The worst-case fidelity is
\begin{align}\label{eq:CorreworstFidelity}
F_{\min}^{2}=1-(2+2(\sqrt{2}-1)^2)|g|^4 t^2+O(t^3).
\end{align}
\end{theorem}
This theorem will be proved in Section \ref{sec:CorrWorstFideltiy}.

\section{A convenient form for quantum conditional mutual information}\label{sec:QuanMutualInfo}
In this section, we derive a convenient expression for quantum conditional mutual information (QCMI), defined in \eqref{def:Mutual}. The main result is Lemma \ref{LM:mutual} below.

We use $\rho_{RSE}$ to denote the density matrix of the whole system:
\begin{align*}
\rho_{RSE}(t)
:={}&\frac{1}{2}e^{-itH}\Bigg(
\ket{R_1}\ket{0_L}\ket{\Omega}
+\ket{R_2}\ket{1_{L}}\ket{\Omega}
\Bigg)\Bigg(
\bra{R_1}\bra{0_L}\bra{\Omega}
+\bra{R_2}\bra{1_L}\bra{\Omega}
\Bigg)e^{itH}
\end{align*} where the Hamiltonian $H$ is defined in \eqref{def:Hamil} for independent errors and in \eqref{def:hami2} for correlated errors. A crucial simplification is that the reference system $R$ is not subject to any external forces.

Two quantities in the definition of QCMI in \eqref{def:Mutual} can be computed exactly.

Since $\rho_{RSE}(t)$ is a pure state,
\begin{align}
    \operatorname{Tr} \rho_{RSE}(t) \log_2 \rho_{RSE}(t) = 0.
\end{align}
The quantity $\operatorname{Tr} \rho_{SE}(t) \log_2 \rho_{SE}(t)$ can also be computed exactly.
Since $R$ is not subject to any external forces, we can write $\rho_{RSE}(t) = \ket{V(t)}\bra{V(t)}$, where
\begin{equation*}
     \ket{V(t)} = \frac{1}{\sqrt{2}} \sum_{k=1,2}  \ket{R_{k}}  \ket{SE_{k}(t)}
\end{equation*}
Here, $\{R_k\}_{k=1,2}$ and $\{SE_{k}(t)\}_{k=1,2}$ are orthonormal sets of vectors for subsystems $R$ and $SE$, respectively.
Thus,
\begin{equation*}
    \rho_{SE}(t) = \operatorname{Tr}_R \rho_{RSE}(t) = \frac{1}{2} \sum_{k=1,2} \ket{SE_k(t)}\bra{SE_k(t)},
\end{equation*}
and, since $\{SE_k(t)\}_{k=1,2}$ is orthonormal,
\begin{equation*}
    \log_2 \rho_{SE}(t) = \left(\log_2 \frac{1}{2}\right) \sum_{k=1,2} \ket{SE_k(t)}\bra{SE_k(t)}.
\end{equation*}
A direct computation then gives
\begin{align}
    \operatorname{Tr} \rho_{SE}(t) \log_2 \rho_{SE}(t) = -1.
\end{align}

Combining the results above, we obtain a simpler expression for $I(R: E\mid S)$:
\begin{lemma}\label{LM:mutual}
\begin{align}
    I(R : E \mid S) = &-\operatorname{Tr} \rho_{RS}(t) \log_2 \rho_{RS}(t) + \operatorname{Tr} \rho_S(t) \log_2 \rho_S(t) + 1\nonumber\\
    =&-\operatorname{Tr} \rho_{RS}(t) \log_2 \rho_{RS}(t) + \operatorname{Tr} \rho_S(t) \log_2 \big(2\rho_S(t)\big), \label{eq:simpleQCMI}
\end{align} where, in the second step, we used the identity $\operatorname{Tr} \rho_{S}(t)=1$.
\end{lemma}

\section{Proof of \eqref{eq:updatedRhoRS}}\label{sec:IndepDensity}
We reformulate the problem as Lemma \ref{LM:density} below, which directly implies the desired result \eqref{eq:updatedRhoRS}.

Before deriving an explicit expression for $e^{-itH}V_{\mathrm{ini}}$, we consider a simple case to illustrate the key steps.
Specifically, we consider a single physical qubit with possible states $\ket{0}$ and $\ket{1}$, as in \eqref{def:GSES}. The qubit is coupled to a heat bath that is initially in the vacuum state $\ket{\Omega}$.

Define the Hamiltonian $H_0$ by
\begin{equation*}
H_0=Id_S\otimes P+\sigma^+\otimes a^*(0)+\sigma^-\otimes a(0)=Id_S\otimes P+\Pi(0),
\end{equation*}
where $P$, $\sigma^\pm$, $a^*(0)$, and $a(0)$ are defined analogously to the corresponding operators in \eqref{def:Hamil}--\eqref{eq:commute} and have the same properties. The last equality defines $\Pi(0)$.

We first make some preparations before studying
$
e^{-iH_0t}\ket{0}\ket{\Omega}
$ and $
e^{-iH_0t}\ket{1}\ket{\Omega}.
$

By Duhamel\textquotesingle s principle,
\begin{align}
\begin{split}
e^{-iH_0t}
={}&e^{-i\widetilde P t}\Big[
1-i\int_0^t e^{i\widetilde P s}
\Pi(0)
e^{-i\widetilde P s}\,ds-\mathrm{R}\Big],
\end{split}
\end{align}
where $\widetilde P:=Id_S\otimes P$, and
the term $\mathrm{R}$ is defined as
\begin{align*}
\mathrm{R}
:={}&\int_0^t\int_0^{s_1}
e^{i\widetilde P s_1}
\Pi(0) e^{-i\widetilde P(s_1-s_2)}
 e^{-itH_0 s_2}\,ds_2\,ds_1.
\end{align*}

This, together with the definition
\begin{align*}
\Pi(s):=\sigma^+\otimes a^*(s)+\sigma^-\otimes a(s)=e^{i\widetilde P s}\Pi(0)e^{-i\widetilde P s},
\end{align*} and iteration of Duhamel's principle,
implies that
\begin{align}\label{eq:durh}
e^{-iH_0t}
={}&e^{-i\widetilde P t}\Big[
1+\sum_{k=1}^{4}(-i)^k\Xi_{k}(t)
+\mathcal{R}_{\geq 5}(t)
\Big],
\end{align}
where $\Xi_k$, $k\geq 1$, are defined recursively as
\begin{equation*}
\Xi_{1}(t):=\int_0^t\Pi(s)\,ds,
\qquad
\Xi_{k}(t):=\int_0^t\Pi(s)\Xi_{k-1}(s)\,ds
\quad\text{for }k\geq 2,
\end{equation*}
and
\begin{equation*}
\mathcal{R}_{\geq 5}(t)
:=\sum_{k=5}^{\infty}(-i)^k\Xi_{k}(t).
\end{equation*}

%%%%%%%%%%%%%%%%%%%%%%%%%%%%

With these preparations complete, we apply the operator $e^{-itH_0}$ to the vectors $\ket{0}\ket{\Omega}$ and $\ket{1}\ket{\Omega}$.

A direct computation shows that, for any $s\in\mathbb{R}$,
\begin{align*}
\Pi(s)\ket{0}\ket{\Omega}=&\left[\sigma^+\otimes a^*(s)+\sigma^-\otimes a(s)\right]
\ket{0}\ket{\Omega}=0,
\end{align*}
because 
\begin{align}\label{eq:rule01}
\sigma^+\ket{0}=\ket{0},\ \text{and}\ a(s)\ket{\Omega}=0.
\end{align}

Thus, for any $t\in\mathbb{R}$,
\begin{align}\label{eq:singleQubit0}
e^{-itH_0}\ket{0}\ket{\Omega}=\ket{0}\ket{\Omega}.
\end{align}

We now compute $e^{-itH_0}\ket{1}\ket{\Omega}$.

A direct computation gives
\begin{align*}
\Pi(s)\ket{1}\ket{\Omega}=&\left[\sigma^+\otimes a^*(s)+\sigma^-\otimes a(s)\right]
\ket{1}\ket{\Omega}
=\ket{0}\,a^*(s)\ket{\Omega},
\end{align*}
because 
\begin{align}\label{eq:rule02}
\sigma^+\ket{1}=\ket{0}\ \text{and}\ \sigma^-\ket{1}=\ket{0}.
\end{align}
Moreover, for any $s_1,s_2\in\mathbb{R}$,
\begin{align*}
&\Pi(s_1)
\Pi(s_2)
\ket{1}\ket{\Omega}=\Pi(s_1)
\ket{0}\,a^*(s_2)\ket{\Omega}=\sigma^-\ket{0}\,a(s_1)a^*(s_2)\ket{\Omega}=\delta(s_1-s_2)\ket{1}\ket{\Omega}.
\end{align*}
Consequently, a direct computation yields
\begin{align}
\begin{split}
e^{-itH_0}\ket{1}\ket{\Omega}
={}&e^{-it\widetilde P}\Big[
\Big(1-\frac{1}{2}t+\frac{1}{8}t^2\Big)\ket{1}\ket{\Omega}
-i\ket{0}\int_0^t a^*(s_1)\Big(1-\frac{1}{2}s_1\Big)\,ds_1\,\ket{\Omega}\\
&\qquad+\mathrm{Remainder}\,\ket{1}\ket{\Omega}\Big].
\end{split}
\end{align}

To analyze the remainder, we iterate the expansion and use the observations above to obtain
\begin{align*}
\mathrm{Remainder}\,\ket{1}\ket{\Omega}
={}&\eta_1(t)\ket{1}\ket{\Omega}
-i\ket{0}\int_0^t a^*(s_1)\eta_2(s_1)\,ds_1\,\ket{\Omega},
\end{align*}
where $\eta_1$ and $\eta_2$ are smooth real-valued functions satisfying
\begin{align*}
\eta_1(t)=O(t^3),\ \eta_2(t)=O(t^2)\qquad\text{when }|t|\ll 1.
\end{align*}

Thus, for two scalar functions $g$ and $h$ of the form 
\begin{align}\label{def:gtht}
\begin{split}
    g(t)=&1-\frac{1}{2}t+\frac{1}{8}t^2+\mathcal{O}(t^2),\\
h(t)=&1-\frac{1}{2}t+\mathcal{O}(t^2)
\end{split}
\end{align} we have
\begin{align}\label{eq:singleQubit1}
e^{-itH_0}\ket{1}\ket{\Omega}
={}&e^{-it\widetilde P}\Big[
g(t)\ket{1}\ket{\Omega}-i\ket{0}\int_0^t h(s)a^*(s)
\,ds\,\ket{\Omega}
\Big].
\end{align}

We now return to the four-qubit system.

We use the following identity, which is derived in the same way as \eqref{eq:durh}:
\begin{align*}
e^{-itH_{SE}}
={}&e^{-itP}\Big[
1+\sum_{k=1}^{4}(-i)^k\Xi_k(t)+\widetilde{R}_{\geq 5}(t)\Big],
\end{align*}
where $\Xi_k(t)$ is defined recursively by
\begin{align*}
\Xi_1(t)&:=\int_0^t Q(s)\,ds,\\
\Xi_k(t)&:=\int_0^t Q(s)\Xi_{k-1}(s)\,ds,
\qquad k\geq 2,
\end{align*}
and
\begin{equation*}
\widetilde{R}_{\geq 5}(t):=\sum_{k=5}^{\infty}(-i)^k\Xi_k(t),
\end{equation*}
with the operators $P$ and $Q$ defined by
\begin{align*}
P:=&\sum_k Id_{S}\otimes P_k,\\
Q(s)
:={}&\sum_{k=1}^{4}\left(
\sigma_k^{+}\otimes a_k^*(s)
+\sigma_k^{-}\otimes a_k(s)\right)
=e^{isP}Q(0)e^{-isP}.
\end{align*}

We begin by considering
\begin{align*}
\frac{1}{ \sqrt{2} }e^{-itH}
\ket{R_2}\ket{1_L}\ket{\Omega} 
={}& \frac{1}{ \sqrt{2} }\ket{R_2}
e^{-itH_{SE}}\ket{1_L}\ket{\Omega} \\
={}& \frac{1}{2 }\ket{R_2}
\left(
 e^{-it H_{SE} }\ket{1100}\ket{\Omega}
+  e^{-it H_{SE}}\ket{0011}\ket{\Omega}
\right).
\end{align*}
Using the identities \eqref{eq:singleQubit0} and \eqref{eq:singleQubit1}, we obtain
\begin{align*}
e^{-itH_{SE}}\ket{1100}\ket{\Omega} 
={}& e^{-itP}\Big[
g^2(t) \ket{1100}\ket{\Omega}  + g(t)\ket{0100}\int_0^t h(\sigma) a_4^*(\sigma)\,d\sigma\,\ket{\Omega}
 \\
&\quad + g(t)\ket{1000}\int_0^t h(\sigma) a_3^*(\sigma)\,d\sigma\,\ket{\Omega} \\
&\quad + \ket{0000}
\int_0^t h(\sigma_1) a_4^*(\sigma_1)\,d\sigma_1
\int_0^t h(\sigma_2) a_3^*(\sigma_2)\,d\sigma_2\,\ket{\Omega}\Big].
\end{align*}
and
\begin{align*}
e^{-itH_{SE}}\ket{0011}\ket{\Omega} 
={}& e^{-itP}\Big[
g^2(t)\ket{0011}\ket{\Omega}  + g(t)\ket{0001}\int_0^t h(\sigma) a_2^*(\sigma)\,d\sigma\,\ket{\Omega} \\
&\quad + g(t)\ket{0010}\int_0^t h(\sigma) a_1^*(\sigma)\,d\sigma\,\ket{\Omega}
 \\
&\quad + \ket{0000}
\int_0^t h(\sigma_2) a_2^*(\sigma_2)\,d\sigma_2
\int_0^t h(\sigma_1) a_1^*(\sigma_1)\,d\sigma_1\,\ket{\Omega}\Big].
\end{align*}

For the logical state $\ket{0_L}$,
\begin{align*}
\frac{1}{ \sqrt{2} }e^{-itH}
\ket{R_1}\ket{0_L}\ket{\Omega} 
={}& \frac{1}{ \sqrt{2} }\ket{R_1}
e^{-itH_{SE}}\ket{0_L}\ket{\Omega} \\
={}& \frac{1}{2 }\ket{R_1}
\left(
 \ket{0000}\ket{\Omega}
+  e^{-it H_{SE}}\ket{1111}\ket{\Omega}
\right)
\end{align*} where we used the identity $$e^{-itH_{SE}} \ket{0000}\ket{\Omega}=\ket{0000}\ket{\Omega}.$$
A direct computation gives
\begin{align*}
e^{-itH_{SE}}\ket{1111}\ket{\Omega} 
={}& e^{-itP}\Big[
g^4(t)\ket{1111}\ket{\Omega}+g^3(t) A_1+g^2(t) A_2 +A_{\geq 3}\Big]
\end{align*} where $A_k$, $k=1,2,3,4$, denotes the component with $k$ particles in the radiation field:
\begin{align*}
A_1:=& \ket{0111}\int_0^t h(\sigma) a_4^*(\sigma)\,d\sigma\,\ket{\Omega}  + \ket{1011}\int_0^t h(\sigma) a_3^*(\sigma)\,d\sigma\,\ket{\Omega} \\
&\quad + \ket{1101}\int_0^t h(\sigma) a_2^*(\sigma)\,d\sigma\,\ket{\Omega}  + \ket{1110}\int_0^t h(\sigma) a_1^*(\sigma)\,d\sigma\,\ket{\Omega};
\end{align*}
\begin{align*}
A_2={}&\ket{0011}
\prod_{k=3,4}\left(\int_0^t h(s) a_k^*(s)\,ds\right)\ket{\Omega}
+\ket{1100}
\prod_{k=1,2}\left(\int_0^t h(s) a_k^*(s)\,ds\right)\ket{\Omega}\\
&+\ket{1010}
\prod_{k=1,3}\left(\int_0^t h(s) a_k^*(s)\,ds\right)\ket{\Omega}+\ket{0110}
\prod_{k=1,4}\left(\int_0^t h(s) a_k^*(s)\,ds\right)\ket{\Omega}\\
&+\ket{1001}
\prod_{k=2,3}\left(\int_0^t h(s) a_k^*(s)\,ds\right)\ket{\Omega}
+\ket{0101}
\prod_{k=2,4}\left(\int_0^t h(s) a_k^*(s)\,ds\right)\ket{\Omega},
\end{align*}
and $A_{\geq 3}$ collects all components with at least three particles in the radiation field.

Using the expression for $\rho_{RSE}$, a direct computation yields
\begin{align*}
\rho_{RS}=&\operatorname{Tr}_E\rho_{RSE}\\
={}&\frac{1}{4} \operatorname{Tr}_E
\ket{
V_{c}+\sum_{k=1}^{4}V_k
+ \sum_{j,l=1,2,3,4;\ j<l}V_{j,l}
}
\bra{
V+\sum_{k=1}^{4}V_k+ \sum_{j,l=1,2,3,4;\ j<l}V_{j,l}
}
+O(t^3).
\end{align*}
Here, $\ket{V_c}$ is the component in which the radiation field is in the vacuum state:
\begin{align*}
\ket{V_c}={}&\Big[\ket{R_1}
\left(\ket{0000}+g^4(t)\ket{1111}\right)
+g^2(t)\ket{R_2}
\left(\ket{0011}+\ket{1100}\right)\Big]\ket{\Omega},
\end{align*}
For $k=1,2,3,4$, $\ket{V_k}$ is the component with one radiation-field particle associated with the $k$-th qubit:
\begin{align*}
\ket{V_k}
={}&g(t)\sigma_k^+
\left[
g^2(t)\ket{R_1}\ket{1111}
+\ket{R_2}\left(\ket{1100}+\ket{0011}\right)
\right]
\int_0^t h(s) a_k^*(s)\,ds\,\ket{\Omega},
\end{align*}
For $j,\ell=1,2,3,4$ with $j<\ell$, $\ket{V_{j,\ell}}$ is the component with two radiation-field particles associated with the $j$-th and $\ell$-th qubits:
\begin{align*}
\ket{V_{j,\ell}}
={}&\sigma_j^+\sigma_\ell^+
\left[
g^2(t)\ket{R_1}\ket{1111}
+\ket{R_2}\left(\ket{1100}+\ket{0011}\right)
\right]\\
&\times 
\int_0^t h(s_1) a_j^*(s_1)\,ds_1
\int_0^t h(s_2) a_\ell^*(s_2)\,ds_2\,\ket{\Omega}.
\end{align*}
The $O(t^3)$ term arises from the three- and four-particle states in the radiation field. In the derivation, we used the identities
\begin{equation*}
\sigma_k^+\ket{0000}=0,
\qquad
\sigma_k^+\sigma_k^+
=\sigma_k^-\sigma_k^-=0,
\qquad
\sigma^-\sigma^+\ket{1}
=\ket{1}.
\end{equation*}

Before deriving the expression for $\rho_{RS}$ in \eqref{eq:rhoRS1}, we define several vectors.

The vector $\ket{W_1}$ will be used to define the weight-0 error:
\begin{align*}
\ket{W_1}
:={}&\ket{R_1}\left(\ket{0000}+(1-2t +2t^2)\ket{1111}\right)
+(1-t+\frac{1}{2}t^2)\ket{R_2}\left(\ket{0011}+\ket{1100}\right),
\end{align*}
The following four vectors will be used to define the weight-1 errors:
\begin{align*}
\ket{W_2}
:={}&(1-t)\ket{R_1}\ket{1110}+\ket{R_2}\ket{0010},\\
\ket{W_3}
:={}&(1-t)\ket{R_1}\ket{1101}+\ket{R_2}\ket{0001},
\\
\ket{W_4}
:={}&(1-t)\ket{R_1}\ket{1011}+\ket{R_2}\ket{1000},\\
\ket{W_5}
:={}&(1-t)\ket{R_1}\ket{0111}+\ket{R_2}\ket{0100},
\end{align*}
The vectors $\ket{W_k}$, $6\leq k\leq 11$, will be used to define the weight-2 errors:
\begin{align*}
\ket{W_6}
:={}\ket{R_1}\ket{1100}+\ket{R_2}\ket{0000},
\ \ & \ \ 
\ket{W_7}
:={}\ket{R_1}\ket{0011}+\ket{R_2}\ket{0000},\\
\ket{W_8}
:={}\ket{R_1}\ket{1010},
\ \ & \ \ 
\ket{W_9}
:={}\ket{R_1}\ket{0110},
\\
\ket{W_{10}}
:={}\ket{R_1}\ket{1001},
\ \ & \ \ 
\ket{W_{11}}
:={}\ket{R_1}\ket{0101}.
\end{align*}

A direct calculation gives
\begin{align}\label{eq:rhoRS1}
\begin{split}
\rho_{RS}
={}&\frac{1}{4}\operatorname{Tr}_E\Bigg[
\left(
\ket{V_c}+\sum_{k=1}^{4}\ket{V_k}
+\sum_{j<\ell}\ket{V_{j,\ell}}
\right)
\left(
\bra{V_c}+\sum_{k=1}^{4}\bra{V_k}
+\sum_{j<\ell}\bra{V_{j,\ell}}
\right)
\Bigg]+O(t^3)\\
={}&\frac{1}{4}\Big(\ket{W_1}\bra{W_1}
+t(1-\frac{3}{2}t)\sum_{k=2}^{5}\ket{W_k}\bra{W_k}
+t^2\sum_{k=6}^{11}\ket{W_k}\bra{W_k}\Big)+O(t^3).
\end{split}
\end{align}
Here, recalling the definition of $h$ in \eqref{def:gtht}, we used the identity
\begin{align*}
\int_0^t\int_0^t h(s_1)h(s_2)a(s_1)a^*(s_2)\,ds_1\,ds_2\,\ket{\Omega}
={}&\int_0^t\int_0^t h(s_1)h(s_2)\delta(s_1-s_2)\,ds_1\,ds_2\,\ket{\Omega}\\
={}&\left(t-\frac{1}{2}t^2+O(t^3)\right)\ket{\Omega}.
\end{align*}

There is one complication: $\ket{W_6}$ and $\ket{W_7}$ are not orthogonal to each
other. All other pairs of distinct vectors have the desired orthogonality:
\begin{align}\label{eq:nonOrtho}
\ket{W_k}\perp \ket{W_\ell}
\qquad\text{if }(k,\ell)\neq(6,7)\ \text{or}\ (7,6).
\end{align}

This lack of orthogonality prevents a direct application of the following lemma:
\begin{lemma}\label{LM:orthoLog}
Suppose that $\mathcal{V}_1$ and $\mathcal{V}_2$ are two orthogonal
subspaces of a Hilbert space $\mathcal{H}$, i.e., for any
$\ket{f}\in\mathcal{V}_1$ and $\ket{g}\in\mathcal{V}_2$,
$\langle f\mid g\rangle=0$. Suppose that
$\{\ket{V_k}\}_{k=1}^n\subset\mathcal{V}_1$ and
$\{\ket{W_\ell}\}_{\ell=1}^m\subset\mathcal{V}_2$. Then
\begin{align*}
&\left(\sum_{k=1}^n\ket{V_k}\bra{V_k}
+\sum_{\ell=1}^m\ket{W_\ell}\bra{W_\ell}\right)
\log_2\left(\sum_{k=1}^n\ket{V_k}\bra{V_k}
+\sum_{\ell=1}^m\ket{W_\ell}\bra{W_\ell}\right)\\
={}&\left(\sum_{k=1}^n\ket{V_k}\bra{V_k}\right)
\log_2\left(\sum_{k=1}^n\ket{V_k}\bra{V_k}\right)
+\left(\sum_{\ell=1}^m\ket{W_\ell}\bra{W_\ell}\right)
\log_2\left(\sum_{\ell=1}^m\ket{W_\ell}\bra{W_\ell}\right).
\end{align*}
\end{lemma}
\begin{proof}
Let
$A:=\sum_{k=1}^n\ket{V_k}\bra{V_k}$ and
$B:=\sum_{\ell=1}^m\ket{W_\ell}\bra{W_\ell}$. Since
$\mathcal{V}_1$ and $\mathcal{V}_2$ are orthogonal, in orthonormal bases
for these subspaces, $A+B$ is represented by the block-diagonal matrix
\begin{equation*}
\begin{bmatrix}M_1&0\\0&M_2\end{bmatrix},
\end{equation*}
where $M_1$ and $M_2$ represent $A$ and $B$, respectively. Therefore,
\begin{align*}
\begin{bmatrix}M_1&0\\0&M_2\end{bmatrix}
\log_2\begin{bmatrix}M_1&0\\0&M_2\end{bmatrix}
&=\begin{bmatrix}M_1\log_2 M_1&0\\0&M_2\log_2 M_2\end{bmatrix},
\end{align*}
which proves the identity.
\end{proof}

By Lemma \ref{LM:orthoLog} and \eqref{eq:nonOrtho}, we have
\begin{align}\label{eq:RhoRS}
\begin{split}
\rho_{RS}\log_2\rho_{RS}
={}&\frac{1}{4}\ket{W_1}\bra{W_1}
\log_2\left(\frac{1}{4}\ket{W_1}\bra{W_1}\right)\\
&+\sum_{k=2}^{5}\frac{t}{4}(1-\frac{3}{2}t)\ket{W_k}\bra{W_k}
\log_2\left(\frac{t}{4}(1-\frac{3}{2}t)\ket{W_k}\bra{W_k}\right)\\
&+\frac{t^2}{4}\left(
\ket{W_6}\bra{W_6}+\ket{W_7}\bra{W_7}
\right)
\log_2\left(
\frac{t^2}{4}\left(
\ket{W_6}\bra{W_6}+\ket{W_7}\bra{W_7}
\right)
\right)\\
&+\sum_{k=8}^{11}\frac{t^2}{4}\ket{W_k}\bra{W_k}
\log_2\left(\frac{t^2}{4}\ket{W_k}\bra{W_k}\right)+O(t^3\ln t).
\end{split}
\end{align}

The operators $\ket{W_k}\bra{W_k}$, for
$1\leq k\leq 11$ and $k\neq 6,7$, have rank one, whereas
$
\ket{W_6}\bra{W_6}+\ket{W_7}\bra{W_7}
$
has rank two. To simplify the analysis of the term $
\ket{W_6}\bra{W_6}+\ket{W_7}\bra{W_7}
$, we seek a decomposition into rank-one
operators associated with orthogonal vectors. The difficulty is that, as discussed above,
$
\ket{W_6}\not\perp\ket{W_7}.
$

To overcome this difficulty, we seek two orthonormal
vectors
$
\ket{\widetilde{W}_6},\ket{\widetilde{W}_7}
\in\operatorname{span}\{\ket{W_6},\ket{W_7}\}
$ such that
\begin{equation*}
\ket{W_6}\bra{W_6}+\ket{W_7}\bra{W_7}
=\ket{\widetilde{W}_6}\bra{\widetilde{W}_6}
+\ket{\widetilde{W}_7}\bra{\widetilde{W}_7}.
\end{equation*}

Before stating the result, we define two unit vectors:
\begin{align}\label{def:wW6W7}
\begin{split}
   \ket{\widetilde{W}_6}:= &\frac{1}{\sqrt{6}}\ket{R_1}
\left(
\ket{1100}+\ket{0011}
\right)
+\sqrt{\frac{2}{3}}\ket{R_2}\ket{0000},\\
\ket{\widetilde{W}_7}:=&\frac{1}{\sqrt{2}}\ket{R_1}
\left(
\ket{0011}-\ket{1100}
\right).
\end{split}
\end{align}
We obtain the following result:
\begin{lemma}\label{LM:w6w7}
The two vectors $\ket{\widetilde{W}_6}$ and $\ket{\widetilde{W}_7}$ are orthonormal, and
\begin{align}\label{eq:w6w7}
\ket{W_6}\bra{W_6}+\ket{W_7}\bra{W_7}
=&3\ket{\widetilde{W}_6}\bra{\widetilde{W}_6}
+\ket{\widetilde{W_7}}\bra{\widetilde{W_7}}.
\end{align}

\end{lemma}
The vectors $\ket{\widetilde{W}_6}$ and $\ket{\widetilde{W}_7}$ are obtained by a direct application of Lemma \ref{LM:ChangeBasis}. We omit the proof.

We now update the expression for $\rho_{RS}$. Define a new set of vectors $\{\ket{\widetilde{W}_k}\}_{k=1}^{11}$ by taking $\ket{\widetilde{W}_6}$ and $\ket{\widetilde{W}_7}$ as defined in \eqref{def:wW6W7} and setting $\ket{\widetilde{W}_k}:=\ket{W_k}$ for $k\neq 6,7$.
Unlike the vectors used in \eqref{eq:rhoRS1}, these vectors are mutually orthogonal:
\begin{equation*}
\braket{\widetilde{W}_k|\widetilde{W}_\ell}=0
\qquad\text{when }k\neq\ell.
\end{equation*}

This yields the following lemma.
\begin{lemma}\label{LM:density}
The reduced density matrix $\rho_{RS}$ has the orthogonal decomposition
\begin{align}
\rho_{RS}
={}&\frac{1}{4}\Big[\ket{\widetilde{W}_1}\bra{\widetilde{W}_1}
+t(1-\frac{3}{2}t)\sum_{k=2}^{5}\ket{\widetilde{W}_k}\bra{\widetilde{W}_k}
+3t^2 \ket{\widetilde{W}_6}\bra{\widetilde{W}_6}
+t^2\ket{\widetilde{W}_7}\bra{\widetilde{W}_7}\nonumber\\
&\quad +t^2\sum_{\ell=8}^{11}\ket{\widetilde{W}_\ell}\bra{\widetilde{W}_\ell}+O(t^3)\Big]\nonumber
\end{align}

\end{lemma}

This expression for $\rho_{RS}$, together with the definitions of its terms, implies the desired result \eqref{eq:updatedRhoRS} and completes the proof.
 
%%%%%%%%%%%%%%%%%%%%%%%%%%%%%%%%%%%%%%%

\section{Proof of \eqref{eq:IndepEntropy} }\label{sec:IndepEntropy}
Because the vectors in the decomposition of $\rho_{RS}$ are mutually orthogonal,
\begin{align*}
\rho_{RS}\log_2\rho_{RS}
={}&
\sum_{k=1}^{6}E_k \log_2E_k\\
&+\frac{1}{4}\sum_{\ell=7}^{11}
\ket{R_1}\ket{S_{\ell,1}} 
\bra{R_1}\bra{S_{\ell,1}}\ 
\log_2\Bigg[
\frac{1}{4}
\ket{R_1}\ket{S_{\ell,1}}\ 
\bra{R_1}\bra{S_{\ell,1}}
\Bigg]+O(t^3\ln t),
\end{align*}
where $E_k$ is defined as
\begin{align*}
E_k:=\frac{1}{4}\Big(
\ket{R_1}\ket{S_{k,1}}+\ket{R_2}\ket{S_{k,2}}
\Big)
\Big(
\bra{R_1}\bra{S_{k,1}}+\bra{R_2}\bra{S_{k,2}}
\Big).
\end{align*}

By Lemma \ref{LM:Largest}, we have
the following inequality for $k=1,2,3,4,5,6$:
\begin{align*}
&\operatorname{Tr}_{RS}
(E_k \log_2 E_k)\\
\leq{}&\frac{1}{4}\operatorname{Tr}_S\left[
\ket{S_{k,1}}\bra{S_{k,1}}
\log_2\left(\frac{1}{2}\ket{S_{k,1}}\bra{S_{k,1}}\right)
\right]+\frac{1}{4}\operatorname{Tr}_S\left[
\ket{S_{k,2}}\bra{S_{k,2}}
\log_2\left(\frac{1}{2}\ket{S_{k,2}}\bra{S_{k,2}}\right)
\right].
\end{align*}

Combining the results above, we obtain
\begin{align}\label{eq:LogRhoRS}
\begin{split}
\operatorname{Tr}_{RS}\left(
\rho_{RS}\log_2\rho_{RS}
\right)
\leq & \frac{1}{4}\sum_{k=1}^6 \sum_{n=1,2}\operatorname{Tr}_S\left[
\ket{S_{k,n}}\bra{S_{k,n}}\log_2(\frac{1}{2}\ket{S_{k,n}}\bra{S_{k,n}})\right]\\
&+\frac{1}{4}\sum_{\ell=7}^{11} \operatorname{Tr}_{S} 
\ket{S_{\ell,1}} 
\bra{S_{\ell,1}}\ 
\log_2\Bigg[
\frac{1}{4}
\ket{S_{\ell,1}} 
\bra{S_{\ell,1}}\Bigg]+O(t^3\ln t).
\end{split}
\end{align}

Having analyzed $\rho_{RS}$, we now turn to $\rho_S$.

Taking the partial trace of $\rho_{RS}$ in \eqref{eq:updatedRhoRS} over $R$, we obtain
\begin{align*}
\rho_{S}=\operatorname{Tr}_R\rho_{RS}
={}&\frac{1}{4}\sum_{k=1}^{11}
\ket{S_{k,1}}\bra{S_{k,1}}
+\frac{1}{4}\sum_{\ell=1}^{6}
\ket{S_{\ell,2}}\bra{S_{\ell,2}}+O(t^3).
\end{align*}
A complication arises because
\begin{align}\label{eq:nonOrt}
\langle S_{1,1}|S_{6,2}\rangle\neq 0
\qquad\text{and}\qquad
\langle S_{6,1}|S_{1,2}\rangle\neq 0.
\end{align}

Thus, when applying Lemma \ref{LM:orthoLog} to $
\rho_S\log_2 (2\rho_S)$, we must group the contributions from each non-orthogonal pair:
\begin{align}\label{eq:LogRhoS}
\begin{split}
\rho_S\log_2 (2\rho_S)
={}&\sum_{\substack{k=1\\k\neq 1,6}}^{11}
\frac{1}{4}\ket{S_{k,1}}\bra{S_{k,1}}
\log_2\left(
\frac{1}{2}\ket{S_{k,1}}\bra{S_{k,1}}
\right)\\
&+\sum_{\substack{\ell=1\\\ell\neq 1,6}}^{6}
\frac{1}{4}\ket{S_{\ell,2}}\bra{S_{\ell,2}}
\log_2\left(
\frac{1}{2}\ket{S_{\ell,2}}\bra{S_{\ell,2}}
\right)\\
&+\frac{1}{4}\left(
\ket{S_{1,1}}\bra{S_{1,1}}
+\ket{S_{6,2}}\bra{S_{6,2}}
\right)\log_2\left[
\frac{1}{2}\left(
\ket{S_{1,1}}\bra{S_{1,1}}
+\ket{S_{6,2}}\bra{S_{6,2}}
\right)
\right]\\
&+\frac{1}{4}\left(
\ket{S_{1,2}}\bra{S_{1,2}}
+\ket{S_{6,1}}\bra{S_{6,1}}
\right)\log_2\left[
\frac{1}{2}\left(
\ket{S_{1,2}}\bra{S_{1,2}}
+\ket{S_{6,1}}\bra{S_{6,1}}
\right)
\right]\\
&\quad+O(t^3\ln t).
\end{split}
\end{align}

Combining \eqref{eq:simpleQCMI}, \eqref{eq:LogRhoRS}, and \eqref{eq:LogRhoS}, we obtain the desired inequality \eqref{eq:IndepEntropy} after canceling the common terms.

\section{Proof of Lemma \ref{LM:justification}}\label{sec:recovery}
To prove that the recovery map is completely positive and
trace-preserving, we consider $16\times16$
density matrices expressed in the basis of the $16$ orthonormal vectors listed in \eqref{def:base1}, \eqref{def:base2}, and \eqref{def:base3}.

Let $M=[m_{k,\ell}]_{k,\ell=1}^{16}$ be a $16\times16$ density matrix.
In this representation, our recovery map takes the form
\begin{align*}
\mathcal{R}(M)
={}&\begin{bmatrix}
m_{1,1}&m_{1,2}\\
\overline{m_{1,2}}&m_{2,2}
\end{bmatrix}
+\begin{bmatrix}
0&0\\
0&m_{3,3}
\end{bmatrix}+\sum_{k\in\{4,6,8,10\}}
\begin{bmatrix}
m_{k,k}&m_{k,k+1}\\
\overline{m_{k,k+1}}&m_{k+1,k+1}
\end{bmatrix}
+\sum_{k=12}^{16}
\begin{bmatrix}
m_{k,k}&0\\
0&0
\end{bmatrix}.
\end{align*}

To illustrate why this map is completely positive and trace-preserving,
we consider a simple example involving $3\times3$ density matrices.
Suppose that
\begin{equation*}
\rho=[\rho_{k,j}]_{k,j=1}^{3}.
\end{equation*}
We seek a recovery map $\mathcal{R}$ such that
\begin{align}\label{eq:exampleR}
\mathcal{R}(\rho)=
\begin{bmatrix}
\rho_{33}+\rho_{11}&\rho_{12}\\
\overline{\rho_{12}}&\rho_{22}
\end{bmatrix}.
\end{align}

To express $\mathcal{R}$ in terms of Kraus operators, we define $K_1$ and $K_2$ by
\begin{equation*}
K_1=
\begin{pmatrix}
1&0&0\\
0&1&0
\end{pmatrix},
\qquad
K_2=
\begin{pmatrix}
0&0&1\\
0&0&0
\end{pmatrix}.
\end{equation*}
The map $\mathcal{R}$ in \eqref{eq:exampleR} can then be written as
\begin{equation*}
\mathcal{R}(\rho)=K_1\rho K_1^\dagger+K_2\rho K_2^\dagger.
\end{equation*}
A direct computation gives
\begin{align}
K_1^\dagger K_1+K_2^\dagger K_2=I_{3}.
\end{align} 

Hence, $\mathcal{R}$ is completely positive and trace-preserving, as desired.

An analogous Kraus representation and completeness identity can be established in the general case, proving that $\mathcal{R}$ is completely positive and trace-preserving.

\section{Entanglement fidelity for independent errors: Proof of Theorem \ref{thm:entangleFidelity}}\label{sec:IndEntangleFidelity}
The expression for $\rho_{RS}$ in \eqref{eq:updatedRhoRS} implies that $F_e$ takes the following form:
\begin{align*}
F_e
={}&\frac{1}{8}\sum_{k=1}^{11}B_k+O(t^3),
\end{align*}
where, when $1\leq k\leq 6$,
\begin{align*}
B_k:={}&\bra{0_L}\mathcal{R}\bigl(\ket{S_{k,1}}\bra{S_{k,1}}\bigr)\ket{0_L}
+\bra{0_L}\mathcal{R}\bigl(\ket{S_{k,1}}\bra{S_{k,2}}\bigr)\ket{1_L}\\
&+\bra{1_L}\mathcal{R}\bigl(\ket{S_{k,2}}\bra{S_{k,1}}\bigr)\ket{0_L}
+\bra{1_L}\mathcal{R}\bigl(\ket{S_{k,2}}\bra{S_{k,2}}\bigr)\ket{1_L},
\end{align*}
and when $7\leq k\leq 11$,
\begin{align*}
B_k:={}&\bra{0_L}\mathcal{R}\bigl(\ket{S_{k,1}}\bra{S_{k,1}}\bigr)\ket{0_L}.
\end{align*}

As noted earlier, a complication arises because $\ket{S_{6,1}}\parallel\ket{S_{1,2}}$ and
$\ket{S_{6,2}}\not\perp\ket{S_{1,1}}$. These non-orthogonal pairs contribute to $B_1$ and $B_6$; the remaining pairs of distinct vectors have the desired orthogonality.

We begin by computing $B_1$ and $B_6$, which are the most complicated terms.

To prepare for the application of the recovery map, we use \eqref{def:x1x3}, \eqref{eq:s11Norm}, and \eqref{eq:norS11} to obtain
\begin{align}\label{eq:s1162}
\begin{split}
\ket{S_{1,1}}
&=\sqrt{1+(1-\kappa(t))^2}\left[
\cos\theta_1\,\ket{X_1}+\sin\theta_1\,\ket{X_3}
\right],\\
&=\sqrt{2}\left(1-t+\frac{3 }{2}t^2\right)\ket{X_1}
-\frac{ 1 }{\sqrt{2}}t^2\ket{X_3}+O(t^3),\\
&=\sigma_{1,1}^{(1)}\ket{X_1}
+\sigma_{1,1}^{(2)}\ket{X_3}+O(t^3),
\end{split}
\end{align} where $\kappa(t)$ is defined in \eqref{def:kappat}, and the coefficients $\sigma_{1,1}^{(n)}$, $n=1,2$, are defined by the last equality. By \eqref{def:sk1} and \eqref{def:sk2},
we have
\begin{align}
\ket{S_{1,2}}&=\sqrt{2}(1-t+\frac{1}{2}t^2)\ket{X_2}=\sigma_{1,2}\ket{X_2},
\end{align}
where the last equality defines $\sigma_{1,2}$.

For $B_1$,
the recovery rules \eqref{eq:rules-11}, \eqref{eq:rules-12}, and \eqref{eq:rules-13} give
\begin{align}
B_1=\left(\sigma_{1,1}^{(1)}+\sigma_{1,2} )\right)^2+(\sigma_{1,1}^{(2)})^2+O(t^3)=8(1-2t+3t^2)+O(t^3).
\end{align}
Similarly, for $B_6$, we use \eqref{def:x1x3}, \eqref{eq:zzzz}, and the fact that $\ket{S_{6,1}}\parallel \ket{X_2}$ to obtain
\begin{align}
B_6= t^2+O(t^3).
\end{align}

Next, for $2\leq k\leq 5$, applying the recovery rule
\eqref{eq:r2} gives
\begin{align}
B_k=\left(\sqrt{\braket{S_{k,1}|S_{k,1}}}
+\sqrt{\braket{S_{k,2}|S_{k,2}}}\right)^2=4t-10t^2+O(t^3).
\end{align}

For $7\leq k\leq 11$, applying the recovery rule
\eqref{eq:r3} gives
\begin{align}
\sum_{
k=7}^{11
}B_k=\sum_{k=7}^{11}\braket{S_{k,1}|S_{k,1}}=5t^2.
\end{align}

Summing the preceding expressions gives 
\begin{align*}
F_e=\frac{1}{8}\Big(\sum_{k=1}^{11}B_k\Big)
=\frac{1}{8}\Big( 8(1-2t+3t^2)+4(4t-10t^2)
+t^2+5t^2+O(t^3)\Big)=1-\frac{5}{4}t^2+O(t^3),
\end{align*} which is the desired result in
Theorem \ref{thm:entangleFidelity}.

\section{Worst-case fidelity for independent errors: Proof of Theorem \ref{thm:worstgeneral}}\label{sec:IndWorstFidelity}
We begin by expressing the definition of worst-case fidelity in a convenient form.

Following the derivation of \eqref{eq:rhoRS1}, we obtain
\begin{align*}
\mathcal{A}\left(\ket{\Psi_{\alpha,\beta}}\bra{\Psi_{\alpha,\beta}}\right)
={}&\frac{1}{2}\Bigg(
\ket{K_1}\bra{K_1}
+t\left(1-\frac{3}{2}t\right)
\sum_{m=2}^{5}\ket{K_m}\bra{K_m}
+t^2\sum_{m=6}^{11}\ket{K_m}\bra{K_m}
\Bigg)+O(t^3),
\end{align*}
where
\begin{align*}
\ket{K_1}:={}&
\alpha\left(\ket{0000}+(1-2t+t^2)\ket{1111}\right)+\beta\left(1-t+\frac{t^2}{2}\right)
\left(\ket{1100}+\ket{0011}\right)\\
={}&\alpha\sqrt{1+(1-\kappa(t))^2}\Big[
\cos\theta_1\,\ket{X_1}+\sin\theta_1\,\ket{X_3}
\Big]
+\sqrt{2}\beta(1-t+\frac{t^2}{2})\ket{X_2},
\end{align*}
where, in the second step, we used \eqref{eq:s1162} and the definition
$
\ket{X_2}=\frac{1}{\sqrt{2}}
\left(\ket{1100}+\ket{0011}\right),
$, together with the definition of $\kappa(t)$ in \eqref{def:kappat}. The vectors $X_1$ and $X_3$ are defined in \eqref{def:x1x3}. The remaining vectors are
\begin{align*}
\ket{K_2}:={}&\alpha(1-t)\ket{\widetilde{S}_{2,1}}
+\beta\ket{\widetilde{S}_{2,2}},&\qquad
\ket{K_3}:={}&\alpha(1-t)\ket{\widetilde{S}_{3,1}}
+\beta\ket{\widetilde{S}_{3,2}},\\
\ket{K_4}:={}&\alpha(1-t)\ket{\widetilde{S}_{4,1}}
+\beta\ket{\widetilde{S}_{4,2}},&\qquad
\ket{K_5}:={}&\alpha(1-t)\ket{\widetilde{S}_{5,1}}
+\beta\ket{\widetilde{S}_{5,2}},\\
\ket{K_6}:={}&\alpha\ket{1100}+\beta\ket{0000},&\qquad
\ket{K_7}:={}&\alpha\ket{0011}+\beta\ket{0000},\\
\ket{K_8}:={}&\alpha\ket{\widetilde{S}_{8,1}},&\qquad
\ket{K_9}:={}&\alpha\ket{\widetilde{S}_{9,1}},\\
\ket{K_{10}}:={}&\alpha\ket{\widetilde{S}_{10,1}},&\qquad
\ket{K_{11}}:={}&\alpha\ket{\widetilde{S}_{11,1}}.
\end{align*}

Since the recovery rules are defined with respect to the chosen orthonormal basis, we need to express $\ket{K_6}$ and $\ket{K_7}$ as linear combinations of the basis vectors $\ket{X_k}$, $k=1,2,3$, and $\ket{\widetilde{S}_{7,1}}=\frac{1}{\sqrt{2}}
\left(\ket{0011}-\ket{1100}\right).$

By \eqref{def:x1x3} and \eqref{eq:zzzz},
\begin{align*}
\ket{0000}
={}&\frac{1}{\sqrt{1+(1-\kappa(t))^2}}\Big[
\bigl(\cos\theta-(1-\kappa(t))\sin\theta\bigr)\ket{X_1}+\bigl(\sin\theta+(1-\kappa(t))\cos\theta\bigr)\ket{X_3}
\Big].
\end{align*}
Since $\ket{X_2}=\frac{1}{\sqrt{2}}(\ket{0011}+\ket{1100})$ and $\ket{\widetilde{S}_{7,1}}=\frac{1}{\sqrt{2}}(\ket{0011}-\ket{1100})$,
\begin{align*}
\ket{0011}
&=\frac{1}{\sqrt{2}}
\left(\ket{X_2}+\ket{\widetilde{S}_{7,1}}\right),\\
\ket{1100}
&=\frac{1 }{\sqrt{2}}
\left(\ket{X_2}-\ket{\widetilde{S}_{7,1}}\right).
\end{align*}
Substituting these expressions into the definitions of $\ket{K_6}$ and $\ket{K_7}$,
we obtain
\begin{align*}
\ket{K_6}
={}&\frac{ \alpha}{\sqrt{2}}
\left(\ket{X_2}-\ket{\widetilde{S}_{7,1}}\right)\\
&+\frac{\beta}{\sqrt{1+(1-\kappa(t))^2}}\Big[
\bigl(\cos\theta-(1-\kappa(t))\sin\theta\bigr)\ket{X_1}+\bigl(\sin\theta+(1-\kappa(t))\cos\theta\bigr)\ket{X_3}
\Big]\\
={}&\frac{\alpha}{\sqrt{2}}
\left(\ket{X_2}-\ket{\widetilde{S}_{7,1}}\right)
+\frac{\beta}{\sqrt{2}}\left(\ket{X_1}+\ket{X_3}\right)+O(t),\\
\ket{K_7}
={}&\frac{ \alpha}{\sqrt{2}}
\left(\ket{X_2}+\ket{\widetilde{S}_{7,1}}\right)\\
&+\frac{\beta}{\sqrt{1+(1-\kappa(t))^2}}\Big[
\bigl(\cos\theta-(1-\kappa(t))\sin\theta\bigr)\ket{X_1}+\bigl(\sin\theta+(1-\kappa(t))\cos\theta\bigr)\ket{X_3}
\Big]\\
={}&\frac{\alpha}{\sqrt{2}}
\left(\ket{X_2}+\ket{\widetilde{S}_{7,1}}\right)
+\frac{\beta}{\sqrt{2}}\left(\ket{X_1}+\ket{X_3}\right)+O(t).
\end{align*}

We are now ready to evaluate the worst-case fidelity.

Let $\mathcal{R}$ be the recovery map defined in \eqref{eq:rules-11}-\eqref{eq:r4}. Then
\begin{align}\label{eq:worstCase}
\begin{split}
\left\langle\Psi_{\alpha,\beta}\middle|
\mathcal{R}\left(\mathcal{A}\left(\ket{\Psi_{\alpha,\beta}}\bra{\Psi_{\alpha,\beta}}\right)\right)
\middle|\Psi_{\alpha,\beta}\right\rangle
={}&\frac{1}{2}
\left\langle\Psi_{\alpha,\beta}\middle|
\mathcal{R}\left(\ket{K_1}\bra{K_1}\right)
\middle|\Psi_{\alpha,\beta}\right\rangle\\
&+\frac{1}{2}t\left(1-\frac{3}{2}t\right)
\sum_{m=2}^{5}
\left\langle\Psi_{\alpha,\beta}\middle|
\mathcal{R}\left(\ket{K_m}\bra{K_m}\right)
\middle|\Psi_{\alpha,\beta}\right\rangle\\
&+\frac{t^2}{2}\sum_{m=6}^{11}
\left\langle\Psi_{\alpha,\beta}\middle|
\mathcal{R}\left(\ket{K_m}\bra{K_m}\right)
\middle|\Psi_{\alpha,\beta}\right\rangle+O(t^3).
\end{split}
\end{align}

For $m=2,3,4,5$, by the recovery rule \eqref{eq:r2}, we have
\begin{align*}
\mathcal{R}\left(\ket{K_m}\bra{K_m}\right)
&=\left(\alpha(1-t)\ket{0_L}+\beta\ket{1_L}\right)
\left(\overline{\alpha}(1-t)\bra{0_L}+\overline{\beta}\bra{1_L}\right).
\end{align*} A direct computation gives
\begin{align*}
\bra{\Psi_{\alpha,\beta}}\mathcal{R}\left(\ket{K_m}\bra{K_m}\right)\ket{\Psi_{\alpha,\beta}}
&=\left(|\alpha|^2(1-t)+|\beta|^2\right)^2\\
&=\left(1-|\alpha|^2t\right)^2
=1-2|\alpha|^2t+O(t^2).
\end{align*} Hence,
\begin{align}\label{eq:worst2345}
\begin{split}
t\left(1-\frac{3}{2}t\right)\sum_{m=2}^{5}
\bra{\Psi_{\alpha,\beta}}\mathcal{R}\left(\ket{K_m}\bra{K_m}\right)\ket{\Psi_{\alpha,\beta}}
&=4t\left(1-\left(\frac{3}{2}+2|\alpha|^2\right)t\right)+O(t^3)\\
&=4t-\left(6+8|\alpha|^2\right)t^2+O(t^3).
\end{split}
\end{align}

For $m=8,9,10,11$, by the recovery rule \eqref{eq:r3}, we have
\begin{align}\label{eq:worst891011}
\sum_{m=8}^{11}t^2\left\langle\Psi_{\alpha,\beta}\middle|
\mathcal{R}\left(\ket{K_m}\bra{K_m}\right)
\middle|\Psi_{\alpha,\beta}\right\rangle=\sum_{m=8}^{11}t^2\left\langle\Psi_{\alpha,\beta} \middle| 
|\alpha|^2\ket{0_L}\bra{0_L}
\middle|\Psi_{\alpha,\beta}\right\rangle
=4|\alpha|^4 t^2.
\end{align}

It remains to evaluate the contributions for $m=1,6,7$.

To analyze $\ket{K_1}$, we first recall that
$
\kappa(t)=2t-2t^2
$ from \eqref{def:kappat}.
Thus,
\begin{align*}
\sqrt{1+(1-\kappa(t))^2}=\sqrt{2} (1-t+\frac{3}{2}t^2)+O(t^3)
\end{align*}
Since $\sin\theta=O(t^2)$, the recovery rules \eqref{eq:rules-11}-\eqref{eq:rules-13} give
\begin{align*}
\mathcal{R}\left(\ket{K_1}\bra{K_1}\right)
={}&\mathcal{R}\Big(
\left(c_1\alpha\ket{X_1}+c_2\beta\ket{X_2}\right)
\left(c_1\overline{\alpha}\bra{X_1}+c_2\overline{\beta}\bra{X_2}\right)
\Big)+O(t^3)\\
={}&\left(c_1\alpha\ket{0_L}+c_2\beta\ket{1_L}\right)
\left(c_1\overline{\alpha}\bra{0_L}+c_2\overline{\beta}\bra{1_L}\right)+O(t^3),
\end{align*}
where the coefficients $c_1$ and $c_2$ are defined as follows:
\begin{align*}
c_1:=\sqrt{2}\big(1-t+\frac{3}{2} t^2\big)\ \text{and}\ 
c_2:=\sqrt{2}\big(1-t+\frac{1}{2} t^2\big).
\end{align*}
Thus, since $|\alpha|^2 + |\beta|^2 = 1$,
\begin{align}
\bra{\Psi_{\alpha,\beta}}\mathcal{R}\left(\ket{K_1}\bra{K_1}\right)\ket{\Psi_{\alpha,\beta}}
&=\left(c_1|\alpha|^2+c_2|\beta|^2\right)^2+O(t^3)\nonumber\\
&=2\left(1-2t+ t^2\left(2|\alpha|^2+2\right)\right)+O(t^3).\label{eq:estK1}
\end{align}

To evaluate $\mathcal{R}(\ket{K_6}\bra{K_6})$ up to an $O(t)$ remainder,
we use
\begin{align*}
\ket{K_6}
={}&\frac{1 }{\sqrt{2}}\alpha
\left(\ket{X_2}-\ket{\widetilde{S}_{7,1}}\right)+\frac{\beta}{\sqrt{2}}
\left(\ket{X_1}+\ket{X_3}\right)+O(t)
=\sum_{k=1}^{3}\ket{\eta_k}+O(t),
\end{align*}
where $\ket{\eta_k}$, $k=1,2,3$, are defined as follows:
\begin{align*}
\ket{\eta_1}&:=\frac{1}{\sqrt{2}}
\left(\alpha\ket{X_2}+\beta\ket{X_1}\right),\\
\ket{\eta_2}&:=-\frac{1}{\sqrt{2}}\alpha\ket{\widetilde{S}_{7,1}},\\
\ket{\eta_3}&:=\frac{1}{\sqrt{2}}\beta\ket{X_3}.
\end{align*}

The recovery rules \eqref{eq:rules-11}-\eqref{eq:rules-13}, \eqref{eq:r3}, and \eqref{eq:r4} then yield
\begin{align*}
\mathcal{R}\left(\ket{K_6}\bra{K_6}\right)
={}&\sum_{k=1}^{3}\mathcal{R}\left(\ket{\eta_k}\bra{\eta_k}\right)+O(t)\\
={}&\frac{1}{2}\left(\alpha\ket{1_L}+\beta\ket{0_L}\right)
\left(\overline{\alpha}\bra{1_L}+\overline{\beta}\bra{0_L}\right)+\frac{1}{2}|\alpha|^2\ket{0_L}\bra{0_L}
+\frac{1}{2}|\beta|^2\ket{1_L}\bra{1_L}+O(t).
\end{align*}

An analogous calculation gives
\begin{align*}
\mathcal{R}\left(\ket{K_7}\bra{K_7}\right)
={}&\frac{1}{2}\left(\alpha\ket{1_L}+\beta\ket{0_L}\right)
\left(\overline{\alpha}\bra{1_L}+\overline{\beta}\bra{0_L}\right)+\frac{1}{2}|\alpha|^2\ket{0_L}\bra{0_L}
+\frac{1}{2}|\beta|^2\ket{1_L}\bra{1_L}+O(t).
\end{align*}
Consequently,
\begin{align}\label{eq:worstm67}
\begin{split}
&t^2\sum_{k=6,7}\bra{\Psi_{\alpha,\beta}}\mathcal{R}\left(\ket{K_k}\bra{K_k}\right)\ket{\Psi_{\alpha,\beta}}=t^2\Big(\left|\overline{\alpha}\beta+\overline{\beta}\alpha\right|^2
+|\alpha|^4+|\beta|^4\Big)+O(t^3).
\end{split}
\end{align}

Combining \eqref{eq:worstCase} with the calculations above, we obtain
the expansion
\begin{align}\label{eq:worstgeneral}
\bra{\Psi_{\alpha,\beta}}\mathcal{R}\left(
\mathcal{A}\left(\ket{\Psi_{\alpha,\beta}}\bra{\Psi_{\alpha,\beta}}\right)
\right)\ket{\Psi_{\alpha,\beta}}=1-\frac{1}{2} F(\alpha, \beta) t^2+O(t^3)
\end{align} with $F(\alpha,\beta)$ defined as
$$F(\alpha, \beta):=2 + 4|\alpha|^2 - 5|\alpha|^4 - |\beta|^4 - |\overline{\alpha}\beta + \overline{\beta}\alpha|^2.$$

Under the constraint $|\alpha|^2 + |\beta|^2 = 1$, we claim that
\begin{align}\label{eq:entangleFidelity}
\max_{|\alpha|^2+|\beta|^2=1}F(\alpha, \beta)
= \frac{5}{2}.
\end{align} This implies the desired result in Theorem \ref{thm:worstgeneral}.

We now prove this claim.

First, since $- |\overline{\alpha}\beta + \overline{\beta}\alpha|^2\leq 0$, we have
\[
F(\alpha, \beta) \le 2 + 4|\alpha|^2 - 5|\alpha|^4 - |\beta|^4.
\]
Let $r = |\alpha|^2$. Then $|\beta|^4 = (1-r)^2$, so
\begin{align}\label{eq:52}
F(\alpha, \beta)  &\le  2 + 4r - 5r^2 - (1-r)^2= \frac{5}{2} - 6\left(r - \frac{1}{2}\right)^2 
\le \frac{5}{2}.
\end{align}

The bound is attained because $F(\frac{1}{\sqrt{2}}, i \frac{1}{\sqrt{2}})=\frac{5}{2}$. Together with \eqref{eq:52}, this proves \eqref{eq:entangleFidelity}.

%%%%%%%%%%%%%%%%%%%%%%%%%%%%%%%%%%%%%%%%%%%%%%%%%%%%%%
\section{Proof of \eqref{eq:CorrRho}}\label{sec:CorrProof}
For notational convenience, we write
\begin{equation*}
P:=\sum_k P_k,
\qquad
Q(0):=\sum_k\left(g_k J_-a_k(0)+\overline{g_k} J_+ a_k^*(0)\right),
\qquad
H=P+Q(0)
\end{equation*}

As in the derivation of \eqref{eq:durh}, we obtain
\begin{align*}
e^{-itH}
={}&e^{-itP}\Big(
1+\sum_{k=1}^{\infty}(-i)^k\Upsilon_k(t)\Big)
\end{align*}
where the operators $\Upsilon_{k}$, $k\geq 1$, are defined recursively. For $k=1$,
\begin{align*}
\Upsilon_1(t):=\int_0^t Q(s)\,ds
\end{align*}
and for any $k\geq 2$,
\begin{equation*}
\Upsilon_{k}(t):=\int_0^t Q(s)\,\Upsilon_{k-1}(s)\,ds
\end{equation*}
Here,
for any $s\in\mathbb{R}$, $Q(s)$ is defined by
\begin{equation*}
Q(s):=\sum_k\left[
g_kJ_-a_k(s)+\overline{g_k}J_+a_k^*(s)
\right]:=J_{-} a(s)+J_{+}a^*(s),
\end{equation*}
where the terms in $Q(s)$ are defined in \eqref{def:corrAa}-\eqref{def:sumAa}.

Recall that, from \eqref{def:Leung},
$
\ket{0_L}:=\frac{1}{\sqrt{2}}
\left(\ket{0000}+\ket{1111}\right)$ and $\ket{1_{L}}:=\frac{1}{\sqrt{2}}
\left(\ket{1100}+\ket{0011}\right),
$
and recall the definitions of $\ket{\Psi_k}$, $k=1,2,3$, in \eqref{def:Psi123}.
Define
\begin{align}
\begin{split}\label{def:Omega0121}
\Omega_{0,1}
:={}&\frac{1}{\sqrt{2}}\ket{0000}\ket{\Omega}
+\frac{1}{\sqrt{2}}\left(1-2|g|^2t+2|g|^4t^2\right)
\ket{1111}\ket{\Omega},\\
\Omega_{1,1}
:={}&-i\frac{1}{\sqrt{2}}\ket{\Psi_1}
\int_0^t\Big(1-2|g|^2s-3|g|^2(t-s)\Big)a^*(s)\,ds\,\ket{\Omega},\\
\Omega_{2,1}
:={}& -\frac{1}{\sqrt{2}} \ket{\Psi_3}(\int_{0}^{t} a^*(s)\,ds)^2 \ket{\Omega}.
\end{split}
\end{align} and
\begin{align}
\begin{split}\label{def:Omega1232}
\Omega_{0,2}
:={}&\ket{1_{L}}\ket{\Omega}
-\frac{|g|^2}{\sqrt{2}}t\left(1-\frac{3}{2}|g|^2t\right)
\ket{\Psi_3}\ket{\Omega},\\
\Omega_{1,2}
:={}&-i\frac{1}{\sqrt{2}}\ket{\Psi_2}
\int_0^t\Big(1-3|g|^2s-2|g|^2(t-s)\Big)a^*(s)\,ds\ket{\Omega},\\
\Omega_{2,2}
:=& -\sqrt{2}\ket{0000}(\int_{0}^t a^*(s)\, ds)^2 \ket{\Omega}
\end{split}
\end{align}
The following result is useful:
\begin{lemma}
\begin{align}\label{eq:Correlated1L}
e^{-itH}\ket{0_L}\ket{\Omega}
=e^{-itP}\left[\Omega_{0,1}+\Omega_{1,1}+\Omega_{2,1}+\mathrm{Remainder}\right],
\end{align}
where the terms $\Omega_{0,1}$, $\Omega_{1,1}$, and $\Omega_{2,1}$ are defined in \eqref{def:Omega0121},
and
\begin{align}\label{eq:Correlated0L}
e^{-itH}\ket{1_{L}}\ket{\Omega}
=e^{-itP}\left[\Omega_{0,2}+\Omega_{1,2}+\Omega_{2,2}+\mathrm{Remainder}\right],
\end{align}
where
$\Omega_{k,2}$, $k=0,1,2$, are defined in \eqref{def:Omega1232}.

The remainders in the two equations above are different, but both have a negligible effect on the dynamics at the order considered. Each consists of several terms:
\begin{align*}
\mathrm{Remainder}
={}&O_0(t^3)+O_1(t^{5/2})+O_2(t^2)
+O_{\geq 3}(t^{3/2}).
\end{align*}
Here, $O_n(t^{m/2})$ denotes the contribution from the $n$-particle terms in the radiation field and is
of order $O(t^{m/2})$, where $n=0,1,2,3$ and $m=4,5,6$.
The term $O_{\geq 3}(t^{3/2})$ collects the terms with at least three
particles in the radiation field and is of order $O(t^{3/2})$.

\end{lemma}
This lemma will be proved in Subsections \ref{sub:Correlated0L} and \ref{sub:Correlated1L}.

The proof repeatedly uses the following identities, obtained by direct calculation and repeated application of the operators:
\begin{align}\nonumber
J_+\ket{0_{L}}&=\frac{1}{\sqrt{2}}J_+\ket{1111}=\frac{1}{\sqrt{2}}\ket{\Psi_1}, &\qquad  J_+\ket{1_{L}}&=\frac{1}{\sqrt{2}}\ket{\Psi_2},\\ \nonumber
J_-J_+\ket{0_{L}}&=2\sqrt{2}\ket{1111},&\qquad J_-J_+\ket{1_{L}}&=\sqrt{2}\ket{\Psi_3},\\ \label{eq:JP1111}
J_+J_+\ket{0_{L}}&=\sqrt{2}\ket{\Psi_3},&\qquad J_+J_+\ket{1_{L}}&=2\sqrt{2}\ket{0000},\\ \nonumber
J_+J_-J_+\ket{0_{L}}&=2\sqrt{2}\ket{\Psi_1},&\qquad J_+J_-J_+\ket{1_{L}}&=3\sqrt{2}\ket{\Psi_2},\\ \nonumber
J_-J_+J_+\ket{0_{L}}&=3\sqrt{2}\ket{\Psi_1},&\qquad J_-J_+J_+\ket{1_{L}}&=2\sqrt{2}\ket{\Psi_2},\\ \nonumber
J_-J_+J_-J_+\ket{0_{L}}&=8\sqrt{2}\ket{1111}, &\qquad  J_-J_+J_-J_+\ket{1_{L}}&=6\sqrt{2}\ket{\Psi_3}.\nonumber
\end{align} 

We are ready to prove \eqref{eq:CorrRho}.
\begin{proof}
We focus on the partial trace of the component with one particle in the radiation field. We denote this contribution by $K$:
\begin{align}
K
:={}&\operatorname{Tr}_E\Big[
\big(\ket{R_1}\ket{\Omega_{1,1}}+\ket{R_2}\ket{\Omega_{1,2}}\big)
\big(\bra{\Omega_{1,1}}\bra{R_1}+\bra{\Omega_{1,2}}\bra{R_2}\big)
\Big]\nonumber\\
={}&\sum_{k,l=1,2}\ket{R_k}\operatorname{Tr}_E\left(\ket{\Omega_{1,k}}\bra{\Omega_{1,l}}\right)\bra{R_l}\label{eq:K}.
\end{align}
Now we compute each term.
\begin{align*}
\operatorname{Tr}_E\left(\ket{\Omega_{1,1}}\bra{\Omega_{1,1}}\right)
={}&\frac{|g|^2}{2}\ket{\Psi_1}\bra{\Psi_1}
\int_0^t\int_0^t
h_1(t,s_1)h_1(t,s_2)\delta(s_1-s_2)
\,ds_2\,ds_1\\
={}&\frac{|g|^2}{2}\ket{\Psi_1}\bra{\Psi_1}
\int_0^t h_1^2(t,s)\,ds,\\
={}&\frac{|g|^2}{2}\ket{\Psi_1}\bra{\Psi_1} \Big(t-5|g|^2t^2+\frac{19}{3}|g|^4t^3\Big),
\end{align*}
similarly,
\begin{align*}
\operatorname{Tr}_E\left(\ket{\Omega_{1,2}}\bra{\Omega_{1,2}}\right)
={}&\frac{|g|^2}{2}\ket{\Psi_2}\bra{\Psi_2}
\int_0^t h_2^2(t,s)\,ds,\\
={}&\frac{|g|^2}{2}\ket{\Psi_2}\bra{\Psi_2} \Big(t-5|g|^2t^2+\frac{19}{3}|g|^4t^3\Big),
\end{align*}
\begin{align*}
\operatorname{Tr}_E\left(\ket{\Omega_{1,1}}\bra{\Omega_{1,2}}\right)
={}&\frac{|g|^2}{2}\ket{\Psi_1}\bra{\Psi_2}
\int_0^t h_1(t,s)h_2(t,s)\,ds\\
={}&\frac{|g|^2}{2}\ket{\Psi_1}\bra{\Psi_2}\Big(t-5|g|^2t^2+\frac{37}{6}|g|^4t^3 \Big),
\end{align*}
\begin{align*}
\operatorname{Tr}_E\left(\ket{\Omega_{1,2}}\bra{\Omega_{1,1}}\right)
={}&\frac{|g|^2}{2}\ket{\Psi_2}\bra{\Psi_1} \Big(t-5|g|^2t^2+\frac{37}{6}|g|^4t^3 \Big).
\end{align*} Here, the functions $h_1$ and $h_2$ are defined by
\begin{align*}
h_1(t,s)&:=1-2|g|^2s-3|g|^2(t-s)=1-3|g|^2 t+|g|^2 s,\\
h_2(t,s)&:=1-3|g|^2s-2|g|^2(t-s)=1-2|g|^2 t-|g|^2 s.
\end{align*}
    
Returning to \eqref{eq:K}, we obtain a simple expression for $K$:
\begin{align}
    K=\frac{|g|^2}{2}\Big(t-5|g|^2t^2\Big)\left(\ket{R_1}\ket{\Psi_1}+\ket{R_2}\ket{\Psi_2}\right)\left(\bra{R_1}\bra{\Psi_1}+\bra{R_2}\bra{\Psi_2}\right)+O(t^3).
\end{align}
This corresponds to the second term in \eqref{def:CorrW123}.

The other two terms in \eqref{def:CorrW123} are obtained by direct calculation; we omit the details.

\end{proof}

%%%%%%%%%%%%%

%%%%%%%%%%%%%%%%%%%%%%%%%%%%%%%%%%%%%%%%%%%%%%%%%%%%%%%
\subsection{Proof of \eqref{eq:Correlated0L}}\label{sub:Correlated0L}
Since $a(s)\ket{\Omega}=0$, the first interaction gives
\begin{equation*}
Q(s)\ket{1_{L}}\ket{\Omega}
=
\left(J_+\ket{1_{L}}\right)a^*(s)\ket{\Omega}.
\end{equation*}
Thus, the expansion can also be written as
\begin{align*}
e^{-itH}\ket{1_{L}}\ket{\Omega}
={}&e^{-itP}\Bigg[\ket{1_{L}}\ket{\Omega}
+\sum_{n=1}^{4}(-i)^n \Gamma_n(t)+\text{Remainder}(t)\Bigg].
\end{align*}
Here, the terms $\Gamma_n(t)$ are defined recursively, starting with
\begin{align}\label{eq:gamma1}
\Gamma_1(t)
&:=J_+\ket{1_{L}}\int_0^t a^*(s)\,ds\,\ket{\Omega}=\frac{1}{\sqrt{2}}\ket{\Psi_2}
\int_0^t a^*(s)\,ds\,\ket{\Omega},
\end{align} where, in the last step, we used \eqref{eq:JP1111}.
For any integer $n\geq 2$,
\begin{align*}
\Gamma_n(t) &:=\int_{0}^t Q(s) \Gamma_{n-1}(s)\ ds,
\end{align*} and the term $\text{Remainder}(t)$ is defined as
\begin{align*}
\mathrm{Remainder}(t)
:=\sum_{n=5}^{+\infty}(-i)^{n}\Gamma_n(t).
\end{align*}
For the subsequent analysis, we note that $\Gamma_5$ has no nonzero component in which the radiation field is in the vacuum state. That is,
\begin{equation*}
\Gamma_5(t)\perp \ket{n_1,n_2,n_3,n_4}\ket{\Omega},
\qquad n_k\in\{0,1\},\quad k=1,2,3,4.
\end{equation*}

To analyze $\Gamma_2(t)$, we first compute
\begin{align*}
\widetilde{\Gamma}_2(s_1,s_2):={}& Q(s_1)
\left(J_+\ket{1_{L}}\right)
a^*(s_2)\ket{\Omega}\\
={}&J_+J_+\ket{1_{L}}
a^*(s_1)a^*(s_2)\ket{\Omega}
+J_-J_+\ket{1_{L}}
a(s_1)a^*(s_2)\ket{\Omega}\\
={}&J_+J_+\ket{1_{L}}
a^*(s_1)a^*(s_2)\ket{\Omega}
+J_-J_+\ket{1_{L}}
|g|^2\delta(s_1-s_2)\ket{\Omega}\\
={}&2\sqrt{2}\ket{0000}a^*(s_1)a^*(s_2)\ket{\Omega}
+\sqrt{2}|g|^2\ket{\Psi_3}\delta(s_1-s_2)\ket{\Omega}.
\end{align*}
Integrating over the time-ordered region gives
\begin{align}\label{eq:gamma2}
\Gamma_2(t)
={}&\sqrt{2}\ket{0000}
\left(\int_0^t a^*(s)\,ds\right)^2\ket{\Omega}
+\frac{|g|^2}{\sqrt{2}}t\ket{\Psi_3}\ket{\Omega}.
\end{align}
Here, we used the identity
\begin{align}\label{eq:double}
\int_0^t\int_0^{s_1}a^*(s_1)a^*(s_2)\,ds_2\,ds_1
=\frac{1}{2}\left(\int_0^t a^*(s)\,ds\right)^2.
\end{align}

For $\Gamma_3=\int_0^s Q(s)\Gamma_2(s)\ ds$,
\begin{align*}
Q(s)\Gamma_2(s)
={}&J_-J_+J_+\ket{1_{L}}a(s)
\int_0^s\int_0^{s_1}
a^*(s_1)a^*(s_2)\,ds_2\,ds_1\,\ket{\Omega}+\frac{|g|^2}{2}sJ_+J_-J_+\ket{1_{L}}a^*(s)\ket{\Omega}.
\end{align*}
Here, we used $(J_+)^3\ket{1_{L}}=0$ and $a(s)\ket{\Omega}=0$ to eliminate the vanishing terms.
Integrating this identity and using \eqref{eq:JP1111} gives
\begin{align}\label{eq:gamma3}
\Gamma_3(t)={}&\sqrt{2}|g|^2 \ket{\Psi_2}
\int_0^t (t-s)a^*(s)\,ds\,\ket{\Omega}
+\frac{3\sqrt{2}}{2}|g|^2 \ket{\Psi_2}
\int_0^t s a^*(s)\,ds\,\ket{\Omega},
\end{align} 
where we used the identity
\begin{align}\label{eq:ByParts}
\int_0^t\int_0^{s_1} a^*(s_3)\,ds_3\,ds_1
=\int_0^t (t-s)a^*(s)\,ds
\end{align} obtained by integration by parts, together with
\begin{align}\label{eq:zeroTerm}
\int_0^t\int_0^{s_1}\int_0^{s_2}
a^*(s_3)\,\delta(s_1-s_3)\,ds_3\,ds_2\,ds_1=0.
\end{align}

For $\Gamma_4=\int_0^t Q(s)\Gamma_3(s)\ ds $, 
only the vacuum component contributes to the density matrix at the order considered; all other contributions can be absorbed into the $O(t^3)$ remainder. We denote this component by $\widetilde{\Gamma}_4$:
\begin{align}\label{eq:gamma4}
\widetilde{\Gamma}_4
:={}&\frac{|g|^2}{2}J_-J_+J_-J_+\ket{1_L}
\int_0^t a(s)\int_0^s s_1a^*(s_1)\,ds_1\,ds\,\ket{\Omega}
=t^2\frac{3\sqrt{2}}{4}|g|^2\ket{\Psi_3}\ket{\Omega}.
\end{align}

Combining \eqref{eq:gamma1}, \eqref{eq:gamma2}, \eqref{eq:gamma3}, and \eqref{eq:gamma4}, we obtain the desired result \eqref{eq:Correlated0L}.

\subsection{Proof of \eqref{eq:Correlated1L}}\label{sub:Correlated1L}
By Duhamel's principle,
\begin{align*}
e^{-itH}\ket{0_{L}}\ket{\Omega}
={}&e^{-itP}\Big[\ket{0_{L}}\ket{\Omega}
+\sum_{k=1}^{4}(-i)^k\Phi_k(t)
+\mathrm{Remainder}(t)\Big],
\end{align*}
where $\Phi_1$ is defined as
\begin{align}\label{eq:phi1t}
\Phi_1(t) &:= J_+\ket{0_L}\int_0^t a^*(s)\,ds\,\ket{\Omega}= \frac{1}{\sqrt{2}}\ket{\Psi_1}\int_0^t a^*(s)\,ds\,\ket{\Omega},
\end{align}
where, in the last step, we used \eqref{eq:JP1111}.
For any integer $n\geq 2$, define $\Phi_n$ recursively by
\begin{align*}
\Phi_n(t) &:=\int_{0}^t Q(s) \Phi_{n-1}(s)\ ds,
\end{align*} and
\begin{align*}
\mathrm{Remainder}(t)
&:=\sum_{n=5}^{+\infty}(-i)^{n+1}\Phi_n(t)=O(t^2).
\end{align*}

For the subsequent analysis, we note that $\Phi_5$ has no nonzero component in which the radiation field is in the vacuum state. That is,
\begin{equation*}
\Phi_5(t)\perp \ket{n_1,n_2,n_3,n_4}\ket{\Omega},
\qquad n_k\in\{0,1\},\quad k=1,2,3,4.
\end{equation*}

For $\Phi_2$, 
\begin{align}
\Phi_2(t)
&=\int_0^t\int_0^{s_1}Q(s_1)\left(J_+\ket{0_L}\right)
a^*(s_2)\,ds_2\,ds_1\,\ket{\Omega}\nonumber\\
&=\sqrt{2}\ket{\Psi_3}\int_0^t\int_0^{s_1}
a^*(s_1)a^*(s_2)\,ds_2\,ds_1\,\ket{\Omega}+2\sqrt{2}|g|^2\ket{1111}\int_0^t\int_0^{s_1}
\delta(s_1-s_2)\,ds_2\,ds_1\,\ket{\Omega}\nonumber\\
&=\frac{\sqrt{2}}{2}\ket{\Psi_3}
\left(\int_0^t a^*(s)\,ds\right)^2\ket{\Omega}
+\sqrt{2}|g|^2t\ket{1111}\ket{\Omega}.\label{eq:phi2}
\end{align}
Here, we used the identity \eqref{eq:double}.

For $\Phi_3(t)$,
\begin{align}
Q(s)\Phi_2(s)
={}&J_+J_+J_+\ket{0_{L}}a^*(s)
\int_0^s\int_0^{s_1}a^*(s_1)a^*(s_2)\,ds_2\,ds_1\,\ket{\Omega}\nonumber\\
&+J_-J_+J_+\ket{0_{L}}a(s)
\int_0^s\int_0^{s_1}a^*(s_1)a^*(s_2)\,ds_2\,ds_1\,\ket{\Omega}\nonumber\\
&+\frac{|g|^2}{2}sJ_+J_-J_+\ket{0_{L}}a^*(s)\ket{\Omega}\nonumber\\
={}&J_+J_+J_+\ket{0_{L}}a^*(s)
\int_0^s\int_0^{s_1}a^*(s_1)a^*(s_2)\,ds_2\,ds_1\,\ket{\Omega}\label{eq:ThreeState}\\
&+\frac{3\sqrt{2}}{2} |g|^2 \ket{\Psi_1} 
\int_0^s a^*(s_2)\,ds_2\,\ket{\Omega}+\sqrt{2}|g|^2  \ket{\Psi_1 }s a^*(s)\ket{\Omega}.\nonumber
\end{align}
Here we used that
\begin{align*}
a(s)&\int_0^s\int_0^{s_1}a^*(s_1)a^*(s_2)\,ds_2\,ds_1\,\ket{\Omega}\\
={}&|g|^2\int_0^s\int_0^{s_1}
\delta(s-s_1)a^*(s_2)\,ds_2\,ds_1\,\ket{\Omega}+|g|^2\int_0^s\int_0^{s_1}
a^*(s_1)\delta(s-s_2)\,ds_2\,ds_1\,\ket{\Omega}\\
={}&\frac{|g|^2}{2}\int_0^s a^*(s_2)\,ds_2\,\ket{\Omega},
\end{align*}
and
\begin{equation*}
\int_0^s\int_0^{s_1}a^*(s_1)\delta(s-s_2)\,ds_2\,ds_1=0.
\end{equation*}

The three-particle component, given by the first term in \eqref{eq:ThreeState}, can be absorbed into the $O(t^3)$ remainder of the density matrix $\rho_{SE}$. Thus, the contribution retained at the order considered is
\begin{align}\label{eq:phi3}
\widetilde{\Phi}_3(t)={}&\sqrt{2}|g|^2  \ket{\Psi_1 }
\int_0^t s a^*(s)\,ds\,\ket{\Omega}+\frac{3\sqrt{2}}{2} |g|^2 \ket{\Psi_1}
\int_0^t (t-s) a^*(s)\,ds\,\ket{\Omega}.
\end{align}
Here, we used \eqref{eq:ByParts}.

For $\Phi_4$, only one component contributes to the density
matrix at the order considered; all other contributions can be absorbed into the
$O(t^3)$ remainder. The retained component is the one in which the radiation field is in the vacuum state:
\begin{align}
\widetilde{\Phi}_4(t)
:={}&\frac{|g|^4}{2}J_-J_+J_-J_+\ket{0_L}
\int_0^t a(s)\int_0^s s_1a^*(s_1)\,ds_1\,ds\,\ket{\Omega}\nonumber\\
={}&\sqrt{2}|g|^4t^2\ket{1111}\ket{\Omega}\label{eq:phi4}.
\end{align}
Here we used \eqref{eq:JP1111} and the identity $
\int_0^s s_1\delta(s-s_1)\,ds_1=\frac{s}{2}.
$

Collecting \eqref{eq:phi1t}, \eqref{eq:phi2}, \eqref{eq:phi3}, and \eqref{eq:phi4} gives the desired result in \eqref{eq:Correlated1L}.

\section{Entropy: Proof of \eqref{eq:corrEntropy}}\label{sec:corrEntropy}

Using \eqref{eq:CorrRho}, we now compute the QCMI from the expression in Lemma \ref{LM:mutual}, which involves
$
\rho_{RS}\log_2\rho_{RS}
$ and $
\rho_S\log_2(2\rho_S).
$

Using \eqref{eq:CorrRho} and the favorable orthogonality condition
\eqref{eq:CorOrtho}, we obtain, to the order retained in
\eqref{eq:CorrRho},
\begin{align*}
\rho_{RS}\log_2\rho_{RS}
={}&\frac{1}{2}\sum_{k=1}^{3}
\ket{W_k}\bra{W_k}\log_2\left(
\frac{1}{2}\ket{W_k}\bra{W_k}
\right)+O(t^3\ln t).
\end{align*}

This expression, together with Lemma \ref{LM:Largest} and the definition of $W_k$ in \eqref{def:CorrW123}, implies that
\begin{align}
\operatorname{Tr}_{RS}\left(
\rho_{RS}\log_2\rho_{RS}
\right)
\leq
\frac{1}{2}\sum_{k=1}^{3}\sum_{\ell=1}^{2}
\operatorname{Tr}_S\left[ c_k
\ket{Y_{k,\ell}}\bra{Y_{k,\ell}}
\log_2\left( c_k
\ket{Y_{k,\ell}}\bra{Y_{k,\ell}}
\right)
\right]+O(t^3\ln t),\label{eq:corrLargest}
\end{align}
where the coefficients $c_k$, $k=1,2,3$, are defined by
\begin{align*}
c_1&:=c_3:=0,\\
c_2&:=\frac{1}{2}|g|^2t\left(1-5|g|^2t\right),
\end{align*}
and we used the orthogonality relation
\begin{equation*}
\ket{Y_{k,1}} \perp \ket{Y_{k,2}}
\qquad \text{for every fixed}\ k=1,2,3.
\end{equation*}

On the other hand,
\begin{align*}
\rho_S
=\operatorname{Tr}_R\rho_{RS}
=\frac{1}{2}\sum_{\ell=1}^{2}\sum_{k=1}^{3} c_k
\ket{Y_{k,\ell}}\bra{Y_{k,\ell}}+O(t^3),
\end{align*}
and
\begin{align}\label{eq:corrRhos}
\begin{split}
\operatorname{Tr}_S\left(
\rho_S\log_2(2\rho_S)
\right)
&=\frac{1}{2}\operatorname{Tr}_S\Bigg[
\left(
\ket{Y_{1,1}}\bra{Y_{1,1}}+\ket{Y_{3,2}}\bra{Y_{3,2}}
\right)\log_2\left(
\ket{Y_{1,1}}\bra{Y_{1,1}}+\ket{Y_{3,2}}\bra{Y_{3,2}}
\right)
\Bigg]\\
&\quad+\frac{1}{2}\operatorname{Tr}_S\Bigg[
\left(
\ket{Y_{1,2}}\bra{Y_{1,2}}+\ket{Y_{3,1}}\bra{Y_{3,1}}
\right)\log_2\left(
\ket{Y_{1,2}}\bra{Y_{1,2}}+\ket{Y_{3,1}}\bra{Y_{3,1}}
\right)
\Bigg]\\
&\quad+\frac{1}{2}\operatorname{Tr}_S\left[ c_2
\ket{Y_{2,1}}\bra{Y_{2,1}}\log_2\left( c_2
\ket{Y_{2,1}}\bra{Y_{2,1}}
\right)
\right]\\
&\quad+\frac{1}{2}\operatorname{Tr}_S\left[ c_2
\ket{Y_{2,2}}\bra{Y_{2,2}}\log_2\left( c_2
\ket{Y_{2,2}}\bra{Y_{2,2}}
\right)
\right]+O(t^3\ln t)
\end{split}
\end{align}
where, in the last step, we used the orthogonality relations described in \eqref{eq:CorrNonOrtho}. The only non-orthogonal pairs are
\begin{align*}
\ket{Y_{1,1}}\not\perp\ket{Y_{3,2}},\ \ 
\ket{Y_{1,2}}\not\perp\ket{Y_{3,1}},
\end{align*} All other pairs of distinct vectors are orthogonal.

Combining \eqref{eq:simpleQCMI}, \eqref{eq:corrLargest}, and \eqref{eq:corrRhos}, we obtain the desired inequality \eqref{eq:corrEntropy} after canceling the common terms.

\section{Entanglement fidelity for correlated errors: Proof of Theorem \ref{thm:correlatedEntanglement}}\label{sec:CorrEntangleFidelity}
The analysis is similar to that in Section \ref{sec:IndEntangleFidelity}, so we omit some intermediate steps.

By the definition of entanglement fidelity,
\begin{align*}
F_e:=\bra{\Psi_{RS}}\mathcal{R}(\rho_{RS})\ket{\Psi_{RS}}
={}&\frac{1}{4}\sum_{k=1}^{3}B_k+O(t^3),
\end{align*} where $B_k$, $k=1,2,3$, are defined as 
\begin{align*}
    B_k:=&
\bra{0_L}\mathcal{R}\left(\ket{Y_{k,1}}\bra{Y_{k,1}}\right)\ket{0_L}
+\bra{0_L}\mathcal{R}\left(\ket{Y_{k,1}}\bra{Y_{k,2}}\right)\ket{1_L}\\
&\qquad+\bra{1_L}\mathcal{R}\left(\ket{Y_{k,2}}\bra{Y_{k,1}}\right)\ket{0_L}
+\bra{1_L}\mathcal{R}\left(\ket{Y_{k,2}}\bra{Y_{k,2}}\right)\ket{1_L},
\end{align*}
and
$\mathcal{R}$ is the recovery map defined in \eqref{eq:corrRule1} and \eqref{eq:corrRule2}. Here, $
\ket{\Psi_{RS}}
:=\frac{1}{\sqrt{2}}
\left(\ket{R_1}\ket{0_L}+\ket{R_2}\ket{1_L}\right).
$

The term $B_2$ is straightforward to evaluate because
\begin{align*}
\ket{Y_{2,1}}
&=\| Y_{2,1}\rangle \|\ket{W_{2,1}},\\
\ket{Y_{2,2}}
&=\| Y_{2,2}\rangle \|\ket{W_{2,2}}.
\end{align*}
For notational convenience, define the two vectors \begin{align*}
\ket{S_1}:=\ket{0_L},
\qquad
\ket{S_2}:=\ket{1_L}.
\end{align*}
Using the recovery rule
$\mathcal{R}(\ket{W_{2,n}}\bra{W_{2,m}})=\ket{S_n}\bra{S_m}$ for $n,m=1,2$,
we obtain
\begin{align}
\begin{split}
B_2&=\left(
\| Y_{2,1}\rangle \|
+\| Y_{2,2}\rangle \|
\right)^2\\
&=\frac{1}{2}|g|^2t\left(1-5|g|^2t\right)
\left[\big\|\Psi_1\rangle\big\|+\big\|\Psi_2\rangle\big\|\right]^2+O(t^3)\\
&=8|g|^2t\left(1-5|g|^2t\right)+O(t^3).
\end{split}
\end{align}

Before studying $B_1$ and $B_3$, we need to expand $\ket{Y_{1,1}}$, $\ket{Y_{1,2}}$, $\ket{Y_{3,1}}$, and $\ket{Y_{3,2}}$ in terms of the orthonormal basis $\{\ket{W_{k,\ell}}\}_{k=1,2,3;\ \ell=1,2}$ defined in \eqref{def:CorrW123}.

By \eqref{eq:y11Norm}, \eqref{eq:y1132}, and \eqref{def:Phi1Phi2}, we have
\begin{align}
\begin{bmatrix}
\ket{Y_{1,1}}\\
\ket{Y_{3,2}}
\end{bmatrix}
&=\begin{bmatrix}
\left(1-|g|^2t+\dfrac{3}{2}|g|^4t^2\right)\ket{W_{1,1}}
-2|g|^4t^2\ket{W_{3,2}}+O(t^3)\\[6pt]
\sqrt{2}|g|^2t\left(\ket{W_{1,1}}+\ket{W_{3,2}}\right)+O(t^2)
\end{bmatrix}.\label{eq:CorY1142}
\end{align}

By \eqref{eq:y12Norm}, \eqref{eq:y1231}, and \eqref{def:Phi3Phi4}, we have
\begin{align}
\begin{bmatrix}
\ket{Y_{1,2}}\\
\ket{Y_{3,1}}
\end{bmatrix}&=
\begin{bmatrix}
\left(1-|g|^2t+\dfrac{5}{2}|g|^4t^2\right)\ket{W_{1,2}}
-2\sqrt{2}|g|^4t^2\ket{W_{3,1}}+O(t^3)\\[6pt]
\sqrt{2}|g|^2t\Big[\ket{W_{1,2}}+\sqrt{2}\ket{W_{3,1}}\Big]+O(t^2)
\end{bmatrix}.\label{eq:CorY1241}
\end{align}
A direct computation using the recovery rules gives
\begin{align}
\begin{split}
B_1=&4\left(1-2 |g|^2 t+5|g|^4 t^2 \right)+O(t^3),\\
B_3=&2|g|^4t^2\left(
\sqrt{2}
+1
\right)^2+O(t^3).
\end{split}
\end{align}

Combining the results above, we obtain
\begin{align}
F_e
&=\frac{1}{4}\left(B_1+B_2+B_3\right)+O(t^3)
=1-\left(5-\frac{(1+\sqrt{2})^2}{2}\right)|g|^4t^2+O(t^3).
\end{align} This proves Theorem \ref{thm:correlatedEntanglement}.

%%%%%%%%%%%%%%%%%%%%%%%%%%%%%%%%%%%%%%%%%%%%
\section{Worst-case fidelity for correlated errors: Proof of Theorem \ref{thm:correlatedWorstCase}}\label{sec:CorrWorstFideltiy}
Since the analysis is similar to that in Section \ref{sec:IndWorstFidelity}, we provide only a sketch of the proof.

Consider a normalized logical state of the form
\begin{equation*}
\ket{\Psi_{\alpha,\beta}}:=\alpha\ket{0_L}+\beta\ket{1_L},
\qquad \alpha,\ \beta\in\mathbb{C},\quad
|\alpha|^2+|\beta|^2=1.
\end{equation*}
The fidelity after applying the noise channel and the recovery map is
\begin{align}
\langle \Psi_{\alpha,\beta}|\mathcal{R}\mathcal{A}\left(\ket{\Psi_{\alpha,\beta}}\bra{\Psi_{\alpha,\beta}}\right)|\Psi_{\alpha,\beta}\rangle
=\sum_{m=1}^{3}\langle \Psi_{\alpha,\beta}|\mathcal{R}\left(\ket{K_m}\bra{K_m}\right)|\Psi_{\alpha,\beta}\rangle+O(t^3),
\end{align}
where $\ket{K_m}$, $m=1,2,3$, are defined as
\begin{align*}
\ket{K_1}
&:=\alpha\ket{Y_{1,1}}+\beta\ket{Y_{1,2}}\\
&=\alpha c_1\ket{W_{1,1}}+\alpha c_2\ket{W_{3,2}}+\beta c_3\ket{W_{1,2}}+\beta c_4\ket{W_{3,1}}+O(t^3)\\[0.5em]
\ket{K_2}
&:=\sqrt{\frac{1}{2} |g|^2 t (1-5 |g|^2 t)}(\alpha\ket{Y_{2,1}}+\beta\ket{Y_{2,2}})\\
&=2\sqrt{\frac{1}{2} |g|^2 t (1-5 |g|^2 t)}(\alpha \ket{W_{2,1}}
 +\beta \ket{W_{2,2}},\\[0.5em]
\ket{K_3}
&:=\alpha\ket{Y_{3,1}}+\beta\ket{Y_{3,2}}\\
&= \sqrt{2}|g|^2t\Big(\alpha 
  (\ket{W_{1,2}}+\sqrt{2}\ket{W_{3,1}})+\beta
  (\ket{W_{1,1}}
  +\ket{W_{3,2}})\Big)+O(t^2).
\end{align*} In the expansions above, we used \eqref{eq:CorY1142} and \eqref{eq:CorY1241}, and $c_k\in\mathbb{R}$, $k=1,2,3,4$, are defined as
\begin{align*}
c_1&:=1-|g|^2t+\frac{3}{2}|g|^4t^2, &\qquad
c_2&:=-2|g|^4t^2,\\
c_3&:=1-|g|^2t+\frac{5}{2}|g|^4t^2, &\qquad
c_4&:=-4\sqrt{2}|g|^4t^2,
\end{align*}

Next, we compute the contribution of each term to the fidelity.

For the contribution from $\ket{K_1}$, we have
\begin{align*}
\mathcal{R}\left(\ket{K_1}\bra{K_1}\right)
={}&\mathcal{R}\Big(
\left(\alpha c_1\ket{W_{1,1}}+\beta c_3\ket{W_{1,2}}\right)\left(\overline{\alpha}c_1\bra{W_{1,1}}
+\overline{\beta}c_3\bra{W_{1,2}}\right)\Big)\\
&+\mathcal{R}\Big(
\left(\alpha c_2\ket{W_{3,2}}+\beta c_4\ket{W_{3,1}}\right)\left(\overline{\alpha}c_2\bra{W_{3,2}}
+\overline{\beta}c_4\bra{W_{3,1}}\right)\Big)+O(t^3)\\
={}&\left(\alpha c_1\ket{0_L}+\beta c_3\ket{1_L}\right)
\left(\overline{\alpha}c_1\bra{0_L}
+\overline{\beta}c_3\bra{1_L}\right)+O(t^3).
\end{align*}
Hence, by the recovery rules in \eqref{eq:corrRule1} and \eqref{eq:corrRule2},
\begin{align}
\begin{split}
\bra{\Psi_{\alpha,\beta}}\mathcal{R}\left(\ket{K_1}\bra{K_1}\right)\ket{\Psi_{\alpha,\beta}}
&=\left(c_1|\alpha|^2+c_3|\beta|^2\right)^2\\
&=\left(1-|g|^2t+\frac{3}{2}|g|^4t^2
+|g|^4t^2|\beta|^2\right)^2+O(t^3)\\
&=1-2|g|^2 t+2|g|^4 (2+|\beta|^2) t^2+O(t^3).
\end{split}
\end{align}

The contribution from $\ket{K_2}$ is straightforward to evaluate:
\begin{align}
\begin{split}
&\bra{\Psi_{\alpha,\beta}}\mathcal{R}\left(\ket{K_2}\bra{K_2}\right)\ket{\Psi_{\alpha,\beta}}\\
=& 2|g|^2 t \left(1 - 5|g|^2 t\right) \bra{\Psi_{\alpha,\beta}} \left(\alpha\ket{0_L}+\beta\ket{1_L}\right)\Big(\overline{\alpha}\bra{0_L}
+\overline{\beta}\bra{1_L}\Big) \ket{\Psi_{\alpha,\beta}}\\
=&2|g|^2t\Big(1-5|g|^2t\Big).
\end{split}
\end{align}

For the contribution from $\ket{K_3}$, we have
\begin{align}
\mathcal{R}\left(\ket{K_3}\bra{K_3}\right)
={}&2|g|^4t^2\Big[
\mathcal{R}\Big(\left(\beta\ket{W_{1,1}}+\alpha\ket{W_{1,2}}\right)
\left(\overline{\beta}\bra{W_{1,1}}+\overline{\alpha}\bra{W_{1,2}}\right)\Big)\nonumber\\
&\quad+\mathcal{R}\Big(\left(\sqrt{2}\alpha\ket{W_{3,1}}+\beta\ket{W_{3,2}}\right)
\left(\sqrt{2}\overline{\alpha}\bra{W_{3,1}}+\overline{\beta}\bra{W_{3,2}}\right)\Big)
\Big]+O(t^3)\nonumber\\
={}&2|g|^4t^2\Big[
\left(\beta\ket{0_L}+\alpha\ket{1_L}\right)
\left(\overline{\beta}\bra{0_L}+\overline{\alpha}\bra{1_L}\right)\nonumber\\
&\quad+\left(\sqrt{2}\alpha\ket{0_L}+\beta\ket{1_L}\right)
\left(\sqrt{2}\overline{\alpha}\bra{0_L}+\overline{\beta}\bra{1_L}\right)
\Big]+O(t^3).\nonumber
\end{align}
Thus,
\begin{align}
&\bra{\Psi_{\alpha,\beta}}\mathcal{R}\left(\ket{K_3}\bra{K_3}\right)\ket{\Psi_{\alpha,\beta}}=2|g|^4t^2\Big[
\Big(\overline{\beta}\alpha+\overline{\alpha}\beta\Big)^2
+\Big(\sqrt{2}|\alpha|^2+|\beta|^2\Big)^2
\Big].
\end{align}

Combining the results above, we obtain
\begin{align}
&\sum_{m=1}^{3}\bra{\Psi_{\alpha,\beta}}
\mathcal{R}\left(\ket{K_m}\bra{K_m}\right)\ket{\Psi_{\alpha,\beta}}=1-|g|^4t^2 F(\alpha,\beta)+O(t^3),
\end{align}
where $F$ is defined as
$$F(\alpha, \beta) := 6 - 2|\beta|^2 - 2|\beta\bar{\alpha} + \alpha\bar{\beta}|^2  - 2\left(\sqrt{2}|\alpha|^2 + |\beta|^2\right)^2.$$
We claim that, under the constraint $|\alpha|^2 + |\beta|^2 = 1$,
\begin{align}
F(\alpha, \beta) \leq 2+2(\sqrt{2}-1)^2.
\end{align}
and that the maximum is attained at $\beta = i\frac{1}{\sqrt{2}}$ and $\alpha = \frac{1}{\sqrt{2}}$.
This yields the result in Theorem \ref{thm:correlatedWorstCase}.

To see this, we observe that
\begin{equation*}
F(\alpha,\beta)\leq 6-2|\beta|^2
-2\left(\sqrt{2}|\alpha|^2+|\beta|^2\right)^2.
\end{equation*}
Let $r=|\alpha|^2$, so that $|\beta|^2=1-r$ and $r\in[0,1]$.
Then
\begin{align*}
F(\alpha,\beta)\leq f(r)
:={}&6-2(1-r)-2\left(1+(\sqrt{2}-1)r\right)^2\\
={}&2+2(\sqrt{2}-1)^2
-2(\sqrt{2}-1)^2\left(r-\frac{1}{2}\right)^2\\
\leq{}&2+2(\sqrt{2}-1)^2.
\end{align*}
In the last step, equality holds for $r=\frac{1}{2}$. In particular when
$\alpha=\frac{1}{\sqrt{2}}$ and $\beta=i\frac{1}{\sqrt{2}}$, $$F(\alpha,\beta)=2+2(\sqrt{2}-1)^2.$$
Thus, the maximum is attained and equals $2+2(\sqrt{2}-1)^2$, as desired.

\appendix

\section{Orthogonalization of rank-two self-adjoint operators}\label{app:orthogonalization}
We begin by formulating the problem.

Suppose that $\ket{Z}$ and $\ket{Y}$ are two vectors. We seek an
orthonormal basis $\{\ket{X_1},\ket{X_2}\}\subseteq\operatorname{span}\{\ket{Z},\ket{Y}\}$
such that, in this basis, the self-adjoint operator
$
\ket{Z}\bra{Z}+\ket{Y}\bra{Y}
$
is represented by a diagonal Hermitian matrix.

We begin with some preparations. Without loss of
generality, assume that
\begin{equation*}
\|Z\rangle\|\geq\|\ket{Y}\|.
\end{equation*}

When $\|Z\rangle\|=0$, the problem is trivial.

When $\|Z\rangle\|\not=0, $ define a unit vector $\ket{Z_1}$ by
\begin{equation*}
\ket{Z_1}:=\frac{1}{\|Z\rangle\|}\ket{Z}.
\end{equation*}
For $\ket{Y}$, there exist real numbers $\gamma$, $\alpha$, and $\beta$ and a unit vector $\ket{Z_2}$
such that
\begin{equation*}
\frac{1}{\|Z\rangle\|}\ket{Y}=\frac{1}{\|Z\rangle\|}e^{i\gamma}\ket{\widetilde{Y}}=e^{i\gamma}\left(\alpha \ket{Z_1}+\beta \ket{Z_2}\right),
\qquad \ket{Z_1}\perp \ket{Z_2}
\end{equation*} where the first equality defines $\ket{\widetilde{Y}}$.
Thus,
\begin{align}
\frac{1}{\|Z\rangle\|}
\begin{bmatrix}
\ket{Z}\\
\ket{\widetilde{Y}}
\end{bmatrix}
=
\begin{bmatrix}
1 & 0\\
\alpha & \beta
\end{bmatrix}
\begin{bmatrix}
\ket{Z_1}\\
\ket{Z_2}
\end{bmatrix}.
\end{align}

In the orthonormal basis $\{\ket{Z_1},\ket{Z_2}\}$,
the operator
$
\ket{Z}\bra{Z}+\ket{Y}\bra{Y},
$
is represented by the matrix
\begin{equation*}
M:= \|Z\rangle\|^2
\begin{bmatrix}
1+\alpha^2 & \alpha\beta\\
\alpha\beta & \beta^2
\end{bmatrix}.
\end{equation*}
We seek a new orthonormal basis $\{\ket{X_1},\ket{X_2}\}$ and an angle
$\theta\in[0,2\pi)$ such that
\begin{align}\label{eq:samX1X2}
\begin{bmatrix}
\ket{X_1}\\
\ket{X_2}
\end{bmatrix}
=
\begin{bmatrix}
\cos\theta & -\sin\theta\\
\sin\theta & \cos\theta
\end{bmatrix}
\begin{bmatrix}
\ket{Z_1}\\
\ket{Z_2}
\end{bmatrix},
\end{align}
and, in the new basis
$\{\ket{X_1},\ket{X_2}\}$, the matrix representing $\ket{Z}\bra{Z}+\ket{Y}\bra{Y}$ is diagonal.

The following lemma is used repeatedly throughout the paper. We omit its proof for brevity. Define the constant $\Delta$ by
\begin{equation*}
\Delta:=\sqrt{
\left(1+\alpha^2-\beta^2\right)^2
+4\alpha^2\beta^2
}>0.
\end{equation*} 
\begin{lemma}\label{LM:ChangeBasis}
When $\beta=0$, the problem is trivial.

When $\beta\not=0$, choose $\theta\in [-\frac{\pi}{2}, \frac{\pi}{2})$ to be
\begin{equation*}
\theta=-\frac{1}{2}\arcsin(\frac{2\alpha\beta}{\Delta}).
\end{equation*}
Then the operator $\ket{Z}\bra{Z}+\ket{Y}\bra{Y}$ is diagonal in the new orthonormal basis $\{\ket{X_1},\ket{X_2}\}$ defined by \eqref{eq:samX1X2},
and
\begin{align*}
    \ket{Z}\bra{Z}+\ket{Y}\bra{Y}=&\|Z\rangle\|^2\Big[\dfrac{1+\alpha^2+\beta^2+\Delta}{2} \ket{X_1}\bra{X_1}+\dfrac{1+\alpha^2+\beta^2-\Delta}{2}\ket{X_2}\bra{X_2}\Big].
\end{align*}

\end{lemma}

\section{Measurement of entropy leakage caused by space overlap}\label{app:leakage}
We begin by defining a function $f$ as follows: for any vectors $\ket{V}$ and $\ket{W}$, let
\begin{align}
\begin{split}
    f(\ket{V},\ket{W}):=&\operatorname{Tr}\Big(
\big(\ket{V}\bra{V}+\ket{W}\bra{W}\big)
\log_2\big(
\ket{V}\bra{V}+\ket{W}\bra{W}
\big)
\Big)\\
&\quad -
\operatorname{Tr}\Big(
\ket{V}\bra{V}
\log_2\left(\ket{V}\bra{V}\right)
\Big)-
\operatorname{Tr}\Big(
\ket{W}\bra{W}
\log_2\left(\ket{W}\bra{W}\right)
\Big).
\end{split}
\end{align}
Assume without loss of generality that
$$\|V\rangle\|\geq\|W\rangle\|.$$ 

Before stating the result, we define several vectors and constants.
When $\|V\rangle\|\not=0,$
there exist $\alpha,\beta\in\mathbb{C}$ and a unit vector $\ket{Z}$ such
that
\begin{equation*}
\frac{1}{\|V\rangle\|} \ket{W}=\alpha \ket{V}+\beta \ket{Z},
\qquad
\ket{Z}\perp \ket{V},
\qquad
|\alpha|^2+|\beta|^2
=\frac{\|W\rangle\|^2}{\|V\rangle\|^2}\leq1.
\end{equation*}
Define
\begin{align}
\Delta:=&\sqrt{
\left(1+|\alpha|^2-|\beta|^2\right)^2
+4|\alpha\beta|^2
},\label{def:Delta}\\ 
r:=&\frac{
\Delta-\left(1-|\beta|^2+|\alpha|^2\right)
}{2}\in [0,|\beta|^2).\label{def:r}
\end{align}
We obtain the following result:
\begin{lemma}\label{LM:simplification}
Suppose that $\ket{V}$ and $\ket{W}$ are two vectors.
Then
\begin{align}\label{eq:entropyDe}
f(\ket{V},\ket{W})\geq 0.
\end{align}
Equality holds only when $\ket{V}\perp \ket{W}$.

More precisely,
\begin{itemize}
\item 
If $\|V\rangle\|=0$, then $f(\ket{V},\ket{W})=0$.  

\item If $\|V\rangle\|\not=0$, then 
\begin{align}\label{eq:fvw}
f(\ket{V},\ket{W})=\|V\rangle\|^2\Bigg[{}
&\int_0^{|\alpha|^2}
\log_2\left(
\frac{1+\sigma}{\sigma+|\beta|^2}
\right)d\sigma+\int_0^r
\log_2\left(
\frac{1+|\alpha|^2+s}{|\beta|^2-s}
\right)ds
\Bigg].
\end{align}
\end{itemize}
\end{lemma}

\begin{proof}
We begin by simplifying the problem. It suffices to consider the case
\begin{equation*}
\|V\rangle\|=1.
\end{equation*}
Indeed, for any $\|V\rangle\|\neq0$, define
\begin{equation*}
\ket{\widetilde V}:=\frac{1}{\|V\rangle\|}\ket{V},
\qquad
\ket{\widetilde W}:=\frac{1}{\|V\rangle\|}\ket{W}.
\end{equation*}
Then $\ket{\widetilde V}$ is a unit vector, and
\begin{equation*}
\|\widetilde W \rangle\|^2
=\frac{\|W\rangle\|^2}{\|V\rangle\|^2}
\leq 1.
\end{equation*}
The desired inequality~\eqref{eq:entropyDe} takes the form
\begin{align*}
&\operatorname{Tr}\Bigg[
\|V\rangle\|^2\left(
\ket{\widetilde V}\bra{\widetilde V}
+\ket{\widetilde W}\bra{\widetilde W}
\right)
\log_2\Bigg(
\|V\rangle\|^2\left(
\ket{\widetilde V}\bra{\widetilde V}
+\ket{\widetilde W}\bra{\widetilde W}
\right)
\Bigg)
\Bigg]\\
&\quad\geq
\operatorname{Tr}\Bigg[
\|V\rangle\|^2\ket{\widetilde V}\bra{\widetilde V}
\log_2\left(
\|V\rangle\|^2\ket{\widetilde V}\bra{\widetilde V}
\right)
\Bigg]+
\operatorname{Tr}\Bigg[
\|V\rangle\|^2\ket{\widetilde W}\bra{\widetilde W}
\log_2\left(
\|V\rangle\|^2\ket{\widetilde W}\bra{\widetilde W}
\right)
\Bigg].
\end{align*}
Since the terms involving $\log_2\|V\rangle\|^2$ cancel and $\|V\rangle\|^2>0$,
inequality~\eqref{eq:entropyDe} is equivalent to
\begin{align*}
&\operatorname{Tr}\Bigg[
\left(
\ket{\widetilde V}\bra{\widetilde V}
+\ket{\widetilde W}\bra{\widetilde W}
\right)
\log_2\left(
\ket{\widetilde V}\bra{\widetilde V}
+\ket{\widetilde W}\bra{\widetilde W}
\right)
\Bigg]\\
&\quad\geq
\operatorname{Tr}\left[
\ket{\widetilde V}\bra{\widetilde V}
\log_2\left(
\ket{\widetilde V}\bra{\widetilde V}
\right)
\right]
+\operatorname{Tr}\left[
\ket{\widetilde W}\bra{\widetilde W}
\log_2\left(
\ket{\widetilde W}\bra{\widetilde W}
\right)
\right].
\end{align*}

Thus, it remains to prove~\eqref{eq:entropyDe} under the assumptions
\begin{equation*}
\|W\rangle\|\leq \|V\rangle\|=1.
\end{equation*}

When $\ket{V}\perp \ket{W}$, equality
clearly holds. We next consider the case $\langle V,W\rangle\neq0$.

Since $\|V\rangle\|=1$, there exist $\alpha,\beta\in\mathbb{C}$ and a unit
vector $\ket{Z}$ such that
\begin{align}\label{eq:wform}
\ket{W}=\alpha \ket{V}+\beta \ket{Z},
\qquad
\ket{V}\perp \ket{Z},
\end{align}
and
\begin{equation*}
|\alpha|^2+|\beta|^2=\|W\|^2\leq 1.
\end{equation*}
Thus, in the orthonormal basis $\{V,Z\}$, 
$
\ket{V}\bra{V}+\ket{W}\bra{W}
$
is represented by the matrix
\begin{equation}\label{eq:diagM}
M=
\begin{bmatrix}
1+|\alpha|^2 & \alpha\overline{\beta}\\
\overline{\alpha}\beta & |\beta|^2
\end{bmatrix}.
\end{equation}

By Lemma \ref{LM:ChangeBasis}, there exists a unitary matrix $U$ such that
\begin{equation*}
U^*MU=
\begin{bmatrix}
\lambda_{+} & 0\\[0.75em]
0 & \lambda_{-}
\end{bmatrix},
\end{equation*}
where
\begin{equation*}
\lambda_+:=\frac{1+|\alpha|^2+|\beta|^2+\Delta}{2},
\qquad
\lambda_-:=\frac{1+|\alpha|^2+|\beta|^2-\Delta}{2},
\end{equation*} and $\Delta$ is defined in \eqref{def:Delta}.
The key observation is that
\begin{align}
\lambda_+\geq 1+|\alpha|^2,
\qquad
0\leq\lambda_-\leq |\beta|^2,\ \qquad \lambda_{+}+\lambda_-=1+|\alpha|^2+|\beta|^2.
\end{align}

The reason is the following elementary fact. Any Hermitian matrix of the
form
\begin{equation*}
A=
\begin{bmatrix}
a & b\\
\overline{b} & 0
\end{bmatrix},
\qquad
a\in\mathbb{R},\quad b\in\mathbb{C},
\end{equation*}
has eigenvalues $\mu_+$ and $\mu_-$ satisfying
\begin{equation*}
\mu_+\mu_-=-|b|^2\leq0,
\qquad
\mu_++\mu_-=a.
\end{equation*}
Thus, one eigenvalue is nonnegative and the other is nonpositive; they
have opposite signs when $b\neq0$. Applying this observation to
\begin{equation*}
M-|\beta|^2I
=
\begin{bmatrix}
1+|\alpha|^2-|\beta|^2 & \alpha\overline{\beta}\\
\overline{\alpha}\beta & 0
\end{bmatrix}
\end{equation*}
gives $\lambda_-\leq|\beta|^2$. Since
$\lambda_++\lambda_-=1+|\alpha|^2+|\beta|^2$, it follows that
$\lambda_+\geq1+|\alpha|^2$. Finally, $M$ is positive semidefinite, so
$\lambda_-\geq0$.

A direct computation shows that, for $r$ defined in \eqref{def:r},
\begin{align*}
    \lambda_+=1+|\alpha|^2+r,\ \qquad \lambda_-=|\beta|^2-r.
\end{align*}

Consequently,
\begin{align}\label{eq:vwform}
\begin{split}
&\operatorname{Tr}\Big[
\left(\ket{V}\bra{V}+\ket{W}\bra{W}\right)
\log_2\left(\ket{V}\bra{V}+\ket{W}\bra{W}\right)
\Big]\\
=& (1+|\alpha|^2+r)\log_2(1+|\alpha|^2+r)
+(|\beta|^2-r)\log_2(|\beta|^2-r).
\end{split}
\end{align}
On the other hand, since $\|V\|=1$,
\begin{align}\label{eq:v0}
\operatorname{Tr}\left[
\ket{V}\bra{V}\log_2\left(\ket{V}\bra{V}\right)
\right]=0,
\end{align}
and, by \eqref{eq:wform},
\begin{align}\label{eq:wform2}
&\operatorname{Tr}\left[
\ket{W}\bra{W}\log_2\left(\ket{W}\bra{W}\right)
\right]=\left(|\alpha|^2+|\beta|^2\right)
\log_2\left(|\alpha|^2+|\beta|^2\right).
\end{align}

By the three identities above, proving the desired inequality \eqref{eq:entropyDe} reduces to showing that
\begin{equation*}
G(r)\geq0,\ \text{for}\ r\in[0,|\beta|^2),
\end{equation*}
where $G(r)$ is defined by
\begin{align*}
G(r):={}&
\left(1+|\alpha|^2+r\right)
\log_2\left(1+|\alpha|^2+r\right)+\left(|\beta|^2-r\right)
\log_2\left(|\beta|^2-r\right)\\
&-\left(|\alpha|^2+|\beta|^2\right)
\log_2\left(|\alpha|^2+|\beta|^2\right).
\end{align*}

Observe that
\begin{equation*}
\frac{d}{dr}G(r)
=\log_2\left(1+|\alpha|^2+r\right)
-\log_2\left(|\beta|^2-r\right)>0,
\end{equation*}
because $|\beta|^2\leq1$ and $r\in[0,|\beta|^2]$. Hence,
\begin{align*}
G(r)\geq G(0)
={}&\left(1+|\alpha|^2\right)
\log_2\left(1+|\alpha|^2\right)
+|\beta|^2\log_2|\beta|^2\\
&-\left(|\alpha|^2+|\beta|^2\right)
\log_2\left(|\alpha|^2+|\beta|^2\right)\\
=:{}&\widetilde G\left(|\alpha|^2\right).
\end{align*}
Moreover,
\begin{equation*}
\frac{d}{d|\alpha|^2}\widetilde G\left(|\alpha|^2\right)
=\log_2\left(1+|\alpha|^2\right)
-\log_2\left(|\alpha|^2+|\beta|^2\right)
\geq0.
\end{equation*}
Therefore,
\begin{equation*}
\widetilde G\left(|\alpha|^2\right)
\geq\widetilde G(0)=0.
\end{equation*}
It follows that $G(r)\geq0$, which proves the desired result.

Integrating the derivatives above yields the desired identity \eqref{eq:fvw}.

\end{proof}

%%%%%%%%%%%%%%%%%%%%%%%%%%%%%%%%%%%%%%%%%%%%%%%%%%%%%%%%%%%%%%

\section{Operator convexity}\label{app:convex}

Consider the vector
\begin{equation*}
\ket{W}
=\frac{\alpha}{\sqrt{2}}\ket{R_1}\ket{V_1}
+\frac{\beta}{\sqrt{2}}\ket{R_2}\ket{V_2},
\end{equation*}
where $\ket{V_1}$ and $\ket{V_2}$ are unit vectors, and $\alpha$ and $\beta$ are scalars.

We obtain the following result:
\begin{lemma}\label{LM:Largest}
\begin{align}\label{eq:convexity}
\begin{split}
\operatorname{Tr}_S\Big(\operatorname{Tr}_R \ket{W}\bra{W}\log_2 \big(2\operatorname{Tr}_R \ket{W}\bra{W}\big)\Big)
-\operatorname{Tr}_{RS}\left[
\ket{W}\bra{W}\log_2\left(\ket{W}\bra{W}\right)
\right] \geq0.
\end{split}
\end{align}
\end{lemma}
\begin{proof}

For convenience, we define
\begin{align}\label{def:D1}
D_1:=\frac{1}{2}\operatorname{Tr}_S\left(M\log_2 M\right),
\qquad
M:=2\operatorname{Tr}_R\ket{W}\bra{W},
\end{align}
and
\begin{equation*}
D_2:=\operatorname{Tr}_{RS}\left(
\ket{W}\bra{W}\log_2\left(\ket{W}\bra{W}\right)
\right).
\end{equation*}
Thus, it suffices to prove
\begin{align}\label{eq:inequality}
D_1-D_2\geq0.
\end{align}

We first evaluate $D_2$, which is the simpler term.
Because
$
\frac{2}{|\alpha|^2+|\beta|^2}\ket{W}\bra{W}
$
is a pure state,
\begin{align*}
\ket{W}\bra{W}\log_2\left(\ket{W}\bra{W}\right)
&=\ket{W}\bra{W}
\log_2\left[
\frac{|\alpha|^2+|\beta|^2}{2}
\frac{2}{|\alpha|^2+|\beta|^2}\ket{W}\bra{W}
\right]\\
&=\ket{W}\bra{W}
\log_2\left(\frac{|\alpha|^2+|\beta|^2}{2}\right).
\end{align*}
Therefore,
\begin{align}\label{eq:secondTerm}
D_2=&\operatorname{Tr}_{RS}\left[
\ket{W}\bra{W}\log_2\left(\ket{W}\bra{W}\right)
\right]
=\frac{|\alpha|^2+|\beta|^2}{2}
\log_2\left(\frac{|\alpha|^2+|\beta|^2}{2}\right).
\end{align}

To analyze $D_1$, which is defined in terms of $M$, we write
\begin{align}\label{eq:firstTerm}
\frac{1}{2}M
\log_2\left(M\right)
=\frac{1}{2}F(V_1,V_2),
\end{align}
where
\begin{align*}
F(V_1,V_2)
:={}&\left(
|\alpha|^2\ket{V_1}\bra{V_1}
+|\beta|^2\ket{V_2}\bra{V_2}
\right)
\log_2\left(
|\alpha|^2\ket{V_1}\bra{V_1}
+|\beta|^2\ket{V_2}\bra{V_2}
\right).
\end{align*}
The operators $F(V_1,V_2)$ and $F(V_2,V_1)$ have the same trace, so
\begin{equation*}
\operatorname{Tr}_S F(V_1,V_2)
=\frac{1}{2}\operatorname{Tr}_S\left[
F(V_1,V_2)+F(V_2,V_1)
\right].
\end{equation*}
Since the map $A\mapsto A\log_2A$ is operator convex,
\begin{align*}
&\frac{F(V_1,V_2)+F(V_2,V_1)}{2}\\
\geq&
\frac{|\alpha|^2+|\beta|^2}{2}
\left(\ket{V_1}\bra{V_1}+\ket{V_2}\bra{V_2}\right)
\log_2\left[
\frac{|\alpha|^2+|\beta|^2}{2}
\left(\ket{V_1}\bra{V_1}+\ket{V_2}\bra{V_2}\right)
\right].
\end{align*}

Recall that $\ket{V_1}$ and $\ket{V_2}$ are unit vectors. Then there
exists an orthonormal basis
\begin{equation*}
\{\ket{Z_1},\ket{Z_2}\}
\subset\operatorname{span}\{\ket{V_1},\ket{V_2}\}
\end{equation*}
such that, in this basis, the operator
$\ket{V_1}\bra{V_1}+\ket{V_2}\bra{V_2}$ is represented by a diagonal
Hermitian matrix
\begin{equation*}
H=\begin{bmatrix}
\lambda&0\\
0&2-\lambda
\end{bmatrix},
\qquad 1\leq\lambda\leq2.
\end{equation*}
Then
\begin{align*}
\operatorname{Tr}_S\left[
\frac{F(V_1,V_2)+F(V_2,V_1)}{2}
\right]
&\geq\frac{|\alpha|^2+|\beta|^2}{2}
\left[
\lambda\log_2\lambda+(2-\lambda)\log_2(2-\lambda)
\right]\\
&\quad+(|\alpha|^2+|\beta|^2)
\log_2\left(\frac{|\alpha|^2+|\beta|^2}{2}\right).
\end{align*}

This inequality, together with \eqref{def:D1} and \eqref{eq:firstTerm}, implies that
\begin{align}
D_1\geq{}&\frac{|\alpha|^2+|\beta|^2}{4}
\left[
\lambda\log_2\lambda+(2-\lambda)\log_2(2-\lambda)
\right]+\frac{|\alpha|^2+|\beta|^2}{2}
\log_2\left(\frac{|\alpha|^2+|\beta|^2}{2}\right)\nonumber\\
\geq{}& \frac{|\alpha|^2+|\beta|^2}{2}
\log_2\left(\frac{|\alpha|^2+|\beta|^2}{2}\right).\label{eq:finalD1}
\end{align}
In the second step, we used the fact that, for $1\leq\lambda\leq2$,
\begin{equation*}
\lambda\log_2\lambda+(2-\lambda)\log_2(2-\lambda)\geq0.
\end{equation*}
Indeed, if
\begin{equation*}
f(y):=y\log_2y+(2-y)\log_2(2-y),
\end{equation*}
then $f'(y)\geq0$ for $1\leq y<2$ and $f(1)=0$, with the
convention $0\log_2 0=0$ at the endpoint.

Combining \eqref{eq:secondTerm} and \eqref{eq:finalD1} gives the desired inequality \eqref{eq:inequality}.

\end{proof}

\section*{Acknowledgements}
The author thanks Michael Lawler, Joseph Renes, and Juerg Froehlich.

%\bibliographystyle{abbrv}
%\addcontentsline{toc}{chapter}{Bibliography}
%\bibliography{biblio}
\end{document}